\documentclass[sigplan,9pt,authorversion]{acmart}
\usepackage{listings}
\usepackage{graphbox}
\usepackage{pgfplots}
\usepackage{graphicx}
\usepackage{subcaption}

\newcommand{\syscall}[1]{\texttt{#1}}

\renewcommand\footnotetextcopyrightpermission[1]{} 

\begin{document}

\title{ProvMark: A Provenance Expressiveness Benchmarking System}

\author{Sheung Chi Chan}
\affiliation{
  \institution{University of Edinburgh}
  \country{United Kingdom}
}
\email{s1536869@inf.ed.ac.uk}

\author{James Cheney}
\affiliation{
  \institution{University of Edinburgh}
\institution{The Alan Turing Institute}
  \country{United Kingdom}
}
\email{jcheney@inf.ed.ac.uk}

\author{Pramod Bhatotia}
\affiliation{
  \institution{University of Edinburgh}
\institution{The Alan Turing Institute}
  \country{United Kingdom}
}
\email{pramod.bhatotia@ed.ac.uk}

\author{Thomas Pasquier}
\affiliation{
  \institution{University of Bristol}
  \country{United Kingdom}
}
\email{thomas.pasquier@bristol.ac.uk}

\author{Ashish Gehani}
\affiliation{
  \institution{SRI International}
  \country{United States}
}
\email{ashish.gehani@sri.com}

\author{Hassaan Irshad}
\affiliation{
  \institution{SRI International}
  \country{United States}
}
\email{hassaan.irshad@sri.com}

\author{Lucian Carata}
\affiliation{
  \institution{University of Cambridge}
  \country{United Kingdom}
}
\email{lc525@cam.ac.uk}

\author{Margo Seltzer}
\affiliation{
  \institution{University of British Columbia}
  \country{Canada}
}
\email{mseltzer@cs.ubc.ca}

\begin{abstract}
  System level provenance is of widespread interest for applications
  such as security enforcement and information protection.  However,
  testing the correctness or completeness of provenance capture
  tools is challenging and currently done manually.  In some cases
  there is not even a clear consensus about what behavior is correct.
  We present an automated tool, ProvMark, that uses an existing
  provenance system as a black box and reliably identifies the
  provenance graph structure recorded for a given activity, by a
  reduction to \emph{subgraph isomorphism} problems handled by an
  external solver.  ProvMark is a beginning step in the much needed
  area of testing and comparing the expressiveness of provenance
  systems.  We demonstrate ProvMark's usefuless in comparing three
  capture systems with different architectures and distinct design
  philosophies.
\end{abstract}




\maketitle

\section{Introduction}
Data provenance is information about the origin, history, or
derivation of some information~\cite{moreau10ftws}.  It is commonly
proposed as a basis for reproducibility~\cite{pasquier2017if},
dependability~\cite{alvaro2017abstracting}, and regulatory
compliance~\cite{pasquier2018data}, and it is increasingly being used
for security, through forensic audit~\cite{wang08tissec} or online
dynamic detection of malicious behavior~\cite{han2017frappuccino,
  hassan2018towards}. To cater to the requirements of different
use-cases, there are many system-level provenance capture systems in
the literature, such as PASS~\cite{muniswamy-reddy06usenix},
Hi-Fi~\cite{pohlymmb12}, SPADE~\cite{gehani12middleware},
OPUS~\cite{balakrishnan13tapp}, LPM~\cite{bates15tbm},
Inspector~\cite{inspector}, and CamFlow~\cite{pasquier17socc} covering
a variety of operating systems from Linux and BSD to Android and
Windows. Each system assembles low level system events into a
high-level \emph{provenance graph} describing processes, system
resources, and causal relationships among them.

Often,
such systems are described as capturing a \emph{complete} and
\emph{accurate} description of system activity.  To date, the
designers of each such system have decided how to interpret these
goals independently, making different choices regarding what activity to
record and how to represent it.  Although some of these systems
do use standards such as W3C PROV~\cite{w3c-prov} that establish a
common vocabulary for provenance-related data, such standards
do \emph{not} specify how to record operating system-level behaviour,
or even when such records are considered ``accurate'' or
``complete''.  Indeed, as we shall see, in practice there is little
consensus about how specific activities (e.g., renaming a file) should
be represented in a provenance graph.
Additionally, different systems also work at different
system layers (e.g., kernel space vs. user space), so
some information may be unavailable to a given system.

\begin{figure}[tb]
	\begin{minipage}[b]{0.4\columnwidth}
		\begin{subfigure}[b]{\columnwidth}
			\centering
			\href{https://provmark2018.github.io/sampleResult/spade/rename.svg}{
		    		\includegraphics[scale=0.25]{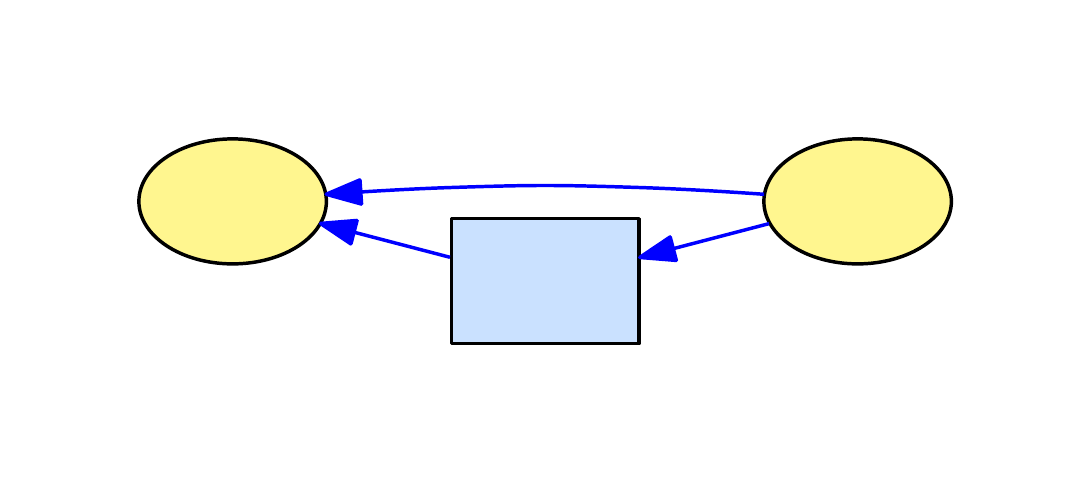}
			}
			\captionof{figure}{SPADE}
		\end{subfigure}
		\\
		\begin{subfigure}[b]{\columnwidth}
			\centering
			\href{https://provmark2018.github.io/sampleResult/camflow/rename.svg}{
		    		\includegraphics[scale=0.25]{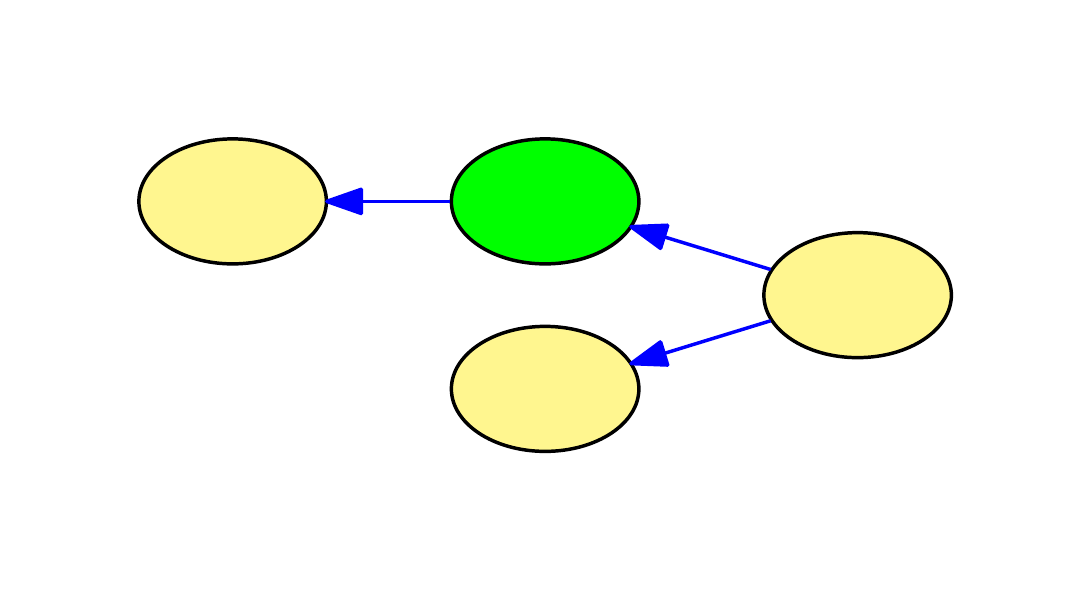}
			}
			\captionof{figure}{CamFlow}
		\end{subfigure}
	\end{minipage}
	\begin{minipage}[b]{0.59\columnwidth}
		\begin{subfigure}[b]{\columnwidth}
			\centering
			\href{https://provmark2018.github.io/sampleResult/opus/rename.svg}{
		    		\includegraphics[scale=0.25]{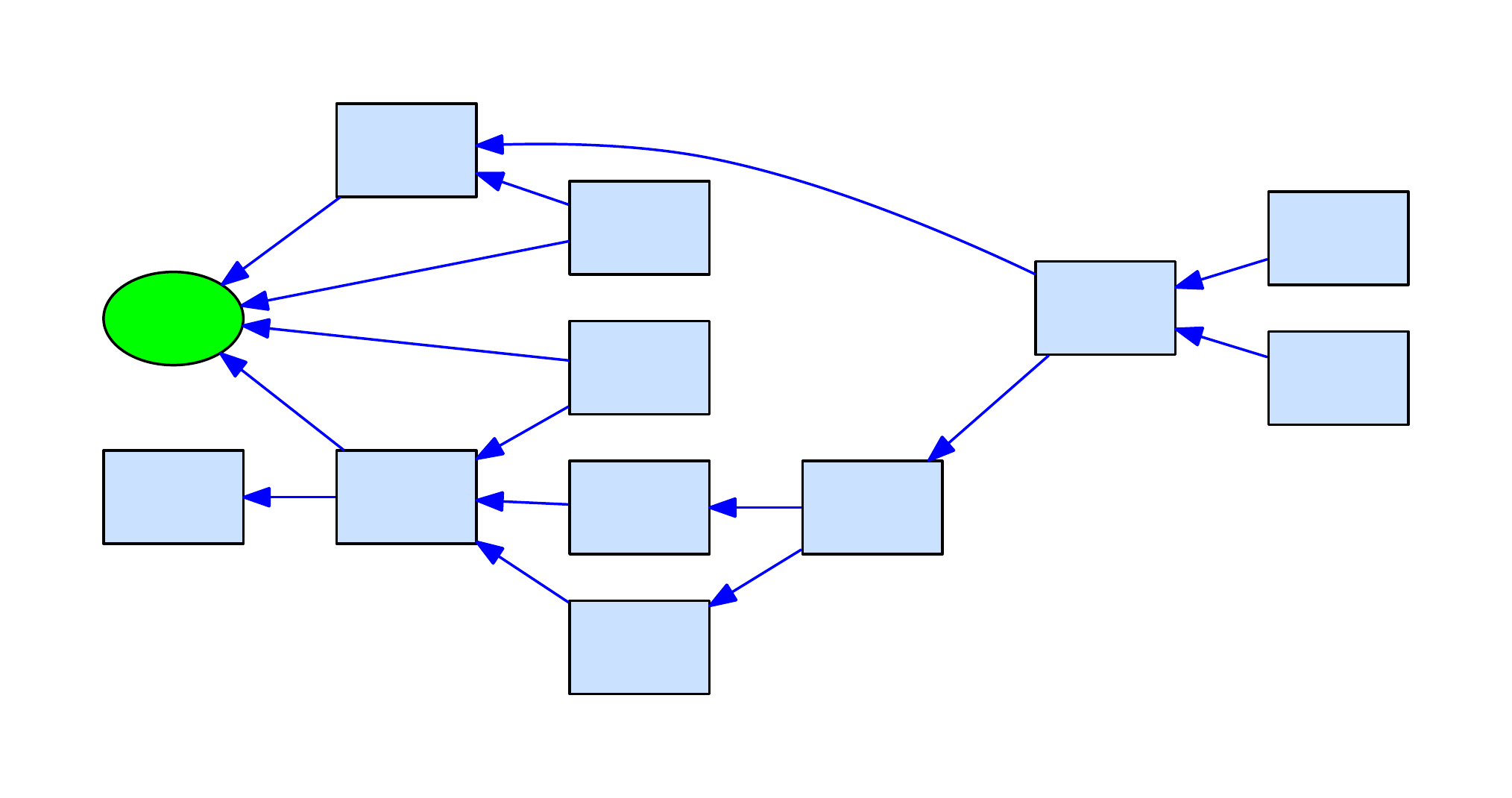}
			}
			\captionof{figure}{OPUS}
		\end{subfigure}
	\end{minipage}	
	\caption{A rename system call, as recorded by three different
    provenance recorders.  Images are clickable links to full-size versions with property labels.}
    \label{fig:rename}
\end{figure}

To illustrate the problem, consider Figure~\ref{fig:rename}, which
shows three provenance graph representations of the same
\syscall{rename} system call, as recorded by three different systems.
These
graphs clearly illustrate nontrivial structural differences in how
\syscall{rename} is represented as communications between processes (blue rectangles) and 
artifacts or resources (yellow ovals).  

Security analysts, auditors, or regulators seeking to use these
systems face many challenges.  Some of these challenges stem from the
inherent size and complexity of provenance graphs: it is not unusual
for a single day's activity to generate graphs with millions of nodes
and edges.  However, if important information is missing, then these
blind spots may render the records useless for their intended purpose.
For example, if a provenance capture system does not record edges
linking reads and writes to local sockets, then attackers can evade
notice by using these communication channels.  Such an omission could
be due to a bug, but it could also result from
misconfiguration, silent failure, or an inherent limitation of the
recording system.

Provenance capture systems provide a
broad-spectrum recording service, separate from the monitored
applications, but (like any software system) they are not perfect, and
can have their own bugs or idiosyncrasies.  In order
to rely on them for critical applications such as reproducible
research, compliance monitoring, or intrusion detection, we need to
be able to understand and validate their behavior.  The strongest form of
validation would consist of verifying that the provenance records
produced by a system are accurate representations of the actual
execution history of the system.  However, while there is now some
work on formalizing operating system kernels, such as seL4~\cite{sel4} and
HyperKernel~\cite{hyperkernel}, there are as yet no complete
formal models of mainstream operating systems such as Linux.
Developing such a model seems prerequisite to fully formalizing the
accuracy, correctness, or completeness of provenance systems.

If there is no immediate prospect of being able to formally define and
prove correctness for provenance recording systems, perhaps we can at
least make it easier to compare and understand their behavior.  We
advocate a pragmatic approach as a first step toward the goal of
validating and testing provenance systems, which (following Chan et
al.~\cite{chan17tapp}) we call \emph{expressiveness benchmarking}.  In
contrast to performance benchmarking, which facilitates quantitative
comparisons of run time or other aspects of a system's behavior,
expressiveness benchmarking facilitates qualitative comparisons of how
different provenance systems record the same activities.  The goal of
such benchmarking is to answer questions such as:

\begin{itemize}
\item What does each node/edge in the graph tell us about actual
  events? (correctness, accuracy)
\item What information is guaranteed to be captured and what are the
  blind spots? (completeness)
\end{itemize}

Previous work on comparing provenance representations has been based
on manual inspection to compare the
graphs~\cite{moreau08challenge,chan17tapp}, but this manual approach
is error prone and does not scale or allow automated testing.
However, automating this task faces numerous challenges.  It is
difficult to configure some systems to record a single process's
activity reproducibly.  Provenance records include volatile
information such as timestamps and identifiers that vary across runs,
even when the underlying process is deterministic.  Furthermore,
a process's activity may include background process start-up activity
that needs to be filtered out.

Expressiveness benchmarking is a  \emph{first (but important) step}
towards understanding what it means for system provenance to be
complete and/or correct.  It offers a systematic means to compare different
approaches.  For example, how is a \syscall{rename} system call
represented, and how do nodes and edges correspond to processes, files
or file versions, or relationships among them?  Are unsuccessful calls
recorded? Are records of multiple similar calls aggregated together?
By analyzing the same benchmark examples across different provenance
capture systems, we
can compare the information collected by different approaches to
produce an understanding of the relative capabilities of the systems
and their suitability for different tasks.  Expressiveness
benchmarking also enables \emph{regression testing} since we can
automatically compare benchmark graphs before and after a system
change to detect differences.

We present \emph{ProvMark}, 
an automated system for provenance
expressiveness benchmarking.
ProvMark executes a given target operation and records the resulting
provenance captured by a given system. Different runs of the same system
can then be compared to show whether each system records the target
activity, and if so how.
Our main contributions are as follows:
\begin{itemize}
\item We show how to generalize from
  multiple runs to obtain repeatable results, abstracting away
  volatile or transient data such as timestamps or identifiers.
\item We solve the required graph matching problems using an external
  solver, based on Answer Set Programming (ASP), a variation of logic
  programming.  
\item We evaluate ProvMark's performance and effectiveness in testing
  and comparing provenance systems, highlighting several bugs or
  idiosyncrasies found so far. 
\end{itemize}

ProvMark has been developed in consultation with developers of three
provenance recording systems, SPADE, OPUS and CamFlow, several of whom
are coauthors of this paper.  They have validated the results
and in some cases helped adapt their systems to ease benchmarking.
The ProvMark system along with supplementary results is publicly
available at \url{http://provmark2018.github.io}~\cite{provmark2018}.

\section{Background and Related Work}
\label{sec:background}

Provenance~\cite{moreau10ftws} is a term originating in the art world.  It refers to the
chain of ownership of a work of art, including its original creator
and each owner and location of the work from creation to the present.
In the digital realm, provenance refers to information describing how
an object came to be in its current form, typically including any
other data from which it was derived and a description of the ways in
which the input data was transformed to produce the output data.  It
is usually collected to allow post-hoc analysis to answer questions
such as, ``From where did this data come?'', ``On what objects does it
depend?'', or ``What programs were used in its production?''
Different applications of provenance will need different data, and
systems are either tailored for specific applications or have a
mechanism for selective capture, so that a user can obtain the right data
for his/her intended use.  Such flexibility makes it difficult to
determine which system or configuration is
most appropriate for a given task.

\begin{figure}[tb]
  \begin{center}
    \includegraphics[scale=0.22]{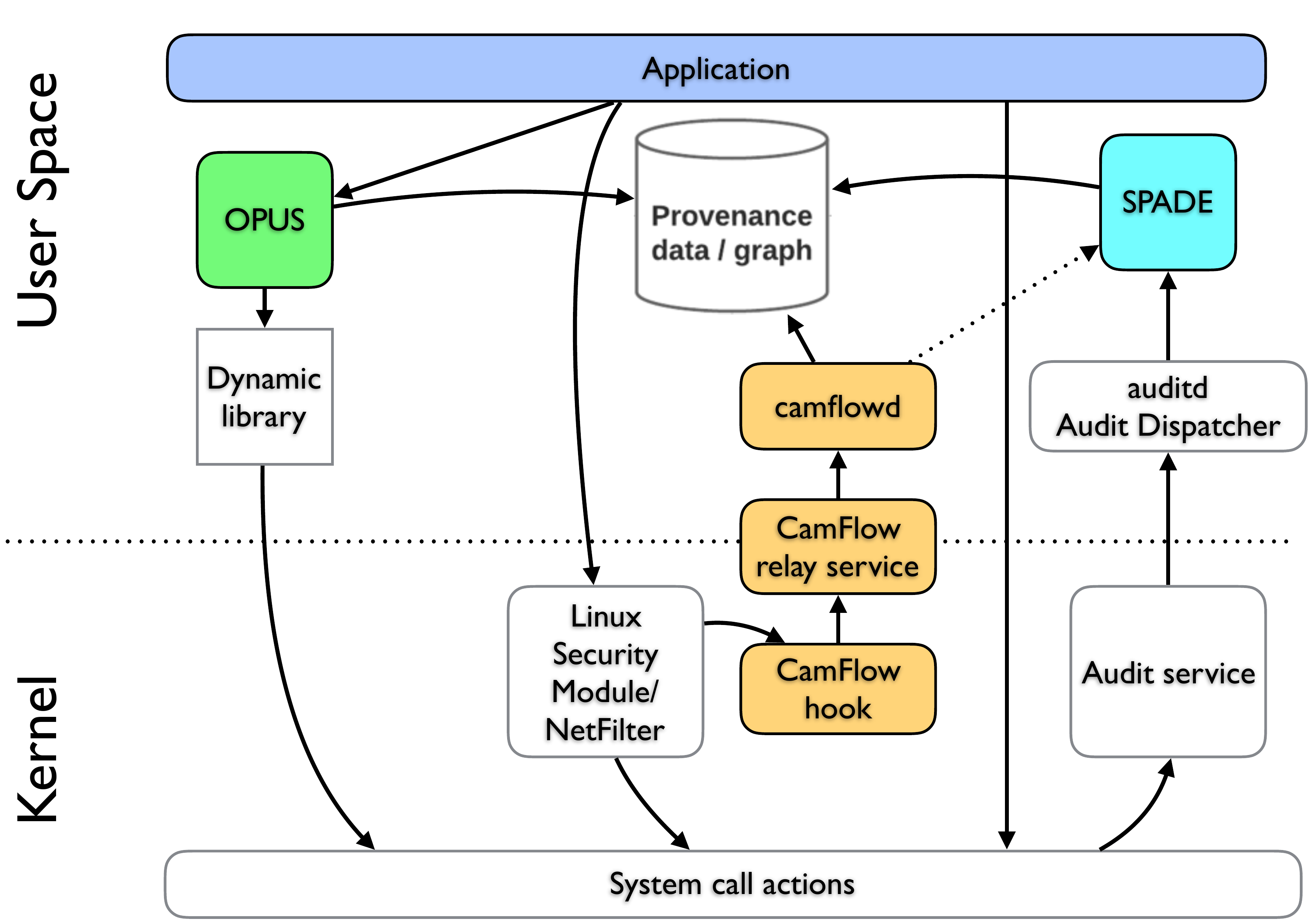}
    \caption{Architecture summary of the SPADE, OPUS, and
      CamFlow provenance recording systems}
\label{graph:recording-architecture}
  \end{center}
\end{figure}

We apply ProvMark to three current provenance capture tools,
namely SPADEv2~\cite{gehani12middleware},
OPUS~\cite{balakrishnan13tapp} and CamFlow~\cite{pasquier17socc}.  All
three systems run under Linux.  There are many other provenance
capture systems.  PASS is no longer
maintained~\cite{muniswamy-reddy06usenix}; HiFi~\cite{pohlymmb12} was
subsumed by LPM~\cite{bates15tbm}, which is similar to CamFlow, but
less portable and not maintained.  We focus on SPADE, OPUS and CamFlow
as representative, currently available examples.
Figure~\ref{graph:recording-architecture} shows how each system
interacts with an application and the Linux kernel.

SPADE's intended use is synthesizing provenance from machines in a
distributed system, so it emphasizes relationships between processes
and digital objects across distributed hosts. Our analysis uses
SPADEv2 (tag \emph{tc-e3}) with the Linux Audit Reporter~\cite{LinuxAudit}, which
constructs a provenance graph using information from the Linux audit
system (including the Audit service, daemon, and dispatcher). SPADE
runs primarily in user space and provides many alternative
configurations, including filtering and transforming the data, for
example, to enable versioning or finer-grained tracking of I/O or
network events, or to make use of information in procfs to obtain
information about processes that were started before SPADE.  We use a
baseline configuration that collects data from the Audit Daemon
(auditd), without versioning.

OPUS focuses on file system
operations, attempting to abstract such operations and make the
provenance collection process portable. It wraps standard C library
calls with hooks that record provenance. The OPUS system is especially
concerned with versioning support and proposes a Provenance Versioning
Model, analogous to models previously introduced in the context of
PASS~\cite{muniswamyreddy09tos} and later SPADE~\cite{gehanitbm11}.
OPUS also runs in user space, but it relies on intercepting calls to
a dynamically linked library (e.g., \verb|libc|). Therefore, it is
blind to activities that do not go through an intercepted dynamic
library, but can observe the C library calls and userspace abstractions
such as file descriptors.

CamFlow's emphasis is sustainability through modularity, interfacing
with the kernel via Linux Security Modules (LSM). The LSM hooks
capture provenance, but then dispatch it to user space, via relayfs,
for further processing.  It strives for completeness and has its roots
in Information Flow Control systems~\cite{pasquier2017camflow}.  By
default, CamFlow captures all system activity visible to LSM and
relates different executions to form a single provenance graph; as we
shall see, this leads to some complications for repeatable
benchmarking.  SPADE can also be configured to support similar
behavior, while CamFlow can also be used (instead of Linux Audit) to
report provenance to SPADE.  Compared to SPADE and OPUS, which both
run primarily in user space, CamFlow~\cite{pasquier17socc} monitors activity and
generates the provenance graph from inside the kernel, via LSM and
NetFilter hooks.  This means the correctness of the provenance data
depends on the LSM operation. As the rules are set directly on the LSM
hooks themselves, which are already intended to monitor all
security-sensitive operations, CamFlow can monitor and/or record all
sensitive operations.  CamFlow allows users to set filtering rules
when collecting provenance.

Prior work~\cite{chan17tapp, moreau08challenge} on comparing different
provenance systems has followed a manual approach.  For example,
the Provenance Challenge~\cite{moreau08challenge} proposed a set of
scientific computation scenarios and solicited submissions
illustrating how different capture systems handle these
scenarios, to facilitate (manual) comparison of systems.  More
recently, Chan et al. proposed a pragmatic, but also manual,
approach~\cite{chan17tapp} to benchmarking OS-level provenance
tracking at the level of individual system calls. However, these
manual approaches are error-prone and not
scalable.
It is also worth mentioning that Pasquier et
al.~\cite{pasquier2018ccs} perform static analysis showing a
conservative over-approximation of the CamFlow provenance recorded for
each call.  However, these results are
not yet automatically compared with actual run-time behavior.

Provenance expressiveness benchmarking is also related to the emerging
topic of \emph{forensic-ready
  systems}~\cite{alrajeh17esecfse,pasquale18icse,zhu15icse}.  Work in
this area considers how to add logging to an existing system to detect
known classes of behaviors, particularly for legal evidence or
regulatory compliance, or how to advise developers on how to add
appropriate logging to systems in development. The provenance
collection systems above strive for completeness so that
previously-unseen behaviors can be detected, and proactively record
information so that user applications do not need to be modified.

\section{System Design and Methodology}
\label{sec:system}

ProvMark is intended to automatically identify the (usually small)
subgraph of a provenance graph that is recorded for a given target
activity.  Target activities could consist of individual system calls
(or \emph{syscalls}), sequences of syscalls, or more general
(e.g., concurrent) processes. For the moment, we consider the simplest
case of a single syscall, but the same techniques generalize to
deterministic sequential target activities; handling concurrency and
nondeterminism are beyond the scope of this paper.

We call the targeted system call the \emph{target call} and the
corresponding subgraph the \emph{target graph}.  Naively, one might
proceed by writing a simple C program for each target system call that
just performs that call and nothing else.  However, starting and
ending a process creates considerable ``boilerplate'' provenance,
including calls such as \syscall{fork}, \syscall{execve}, and \syscall{exit},
as well as accesses to program files and libraries and, sometimes,
memory mapping calls.   Furthermore,
some target calls require other \emph{prerequisite} calls to be performed first.  For
example, analyzing a \syscall{read} or \syscall{close} system call requires first
performing an \syscall{open}.  Process startup and prerequisite calls
are both examples of 
\emph{background activity} that we would like to elide.

In each benchmark, we use an \verb|#ifdef TARGET| CPP directive to
identify the target behavior of interest. ProvMark generates two
executables for each such benchmark: a \emph{foreground program} that
includes all code in the benchmark program, including the target and
any context needed for it to execute, and a \emph{background program}
that contains the background activities. The two binaries are almost
identical; the difference between the resulting graphs should
precisely capture the target behavior.

\begin{figure*}[tb]
  \begin{center}
    \includegraphics[scale=0.4]{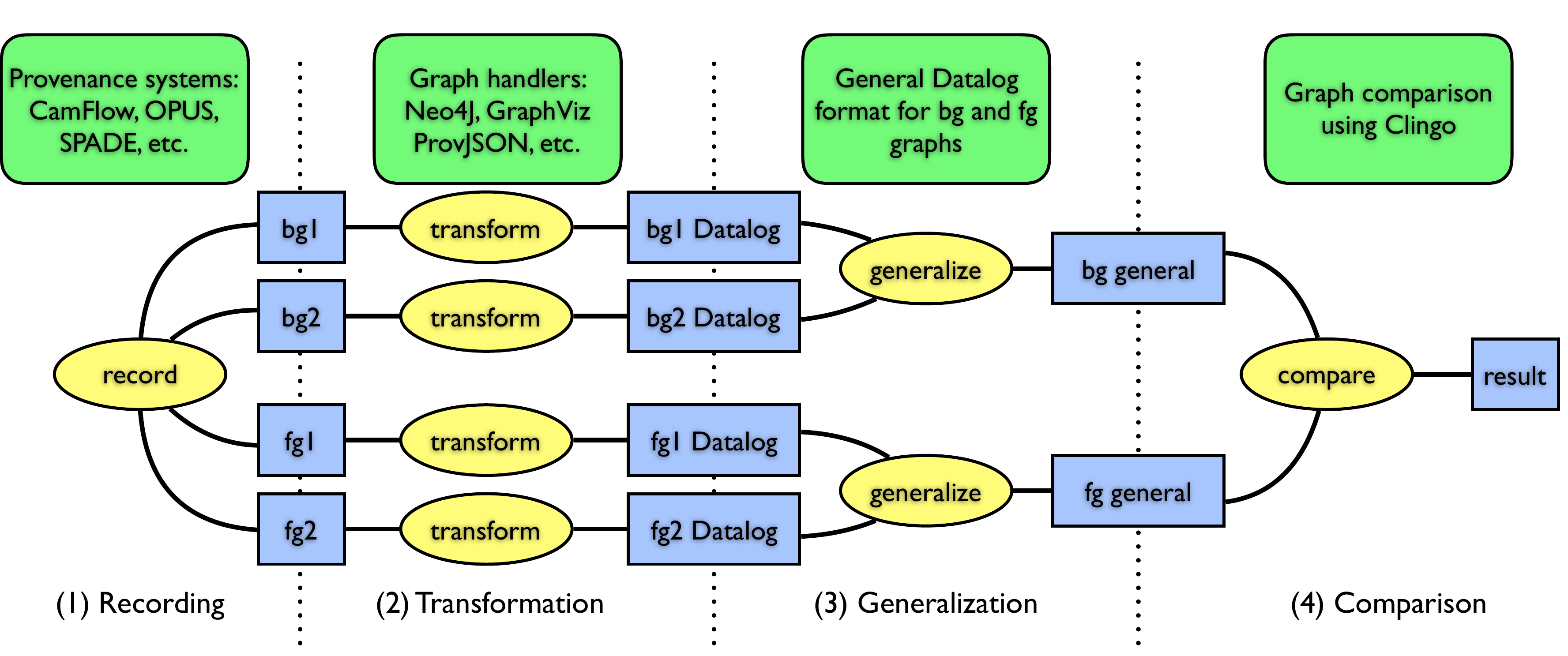}
    \captionof{figure}{ProvMark system overview.  The recording stage (1)
      uses one of the provenance recorders to compute background
      graphs $bg_1,bg_2$ and foreground graphs $fg_1,fg_2$. (The
      same recorder is used for all four graphs.)  The  transformation
    stage (2) maps these graphs to a uniform Datalog format.  The
    generalization stage (3) identifies the common structure of $bg_1$ and
    $bg_2$, resulting in $bg$, and likewise $fg_1$ and $fg_2$ are
    generalized to $fg$.  Finally, $bg$ and $fg$ are compared (4);
    structure corresponding to $bg$ in $fg$ is removed, yielding the
    benchmark result.}
\label{fig:arch}
  \end{center}
\end{figure*}

ProvMark includes a script for each syscall that generates and
compiles the appropriate C executables and prepares a staging
directory in which they will be executed with any needed setup, for
example, first creating a file to run an \syscall{unlink} system
call. We constructed these scripts manually since different
system calls require different setup.  The following code snippet
illustrates the background program (including \syscall{open}) needed
for the target \syscall{close} syscall (with \verb|#ifdef| surrounding
the target):

\footnotesize
\begin{verbatim}
// close.c
#include <fcntl.h>
#include <unistd.h>
void main() {
  int id=open("test.txt", O_RDWR);
#ifdef TARGET 
  close(id); 
#endif
}
\end{verbatim}
\normalsize

Figure~\ref{fig:arch} provides an overview of ProvMark, which is composed of four
subsystems: (1) recording, (2) transformation, (3) generalization, and
(4) comparison.  Users can select which provenance capture system
to use, which benchmark to run, and other configuration settings, such
as the number of trials.  Before presenting the details of the four
subsystems, we outline some common use cases for ProvMark.

\subsection{Use Cases}
\label{sec:usecases}

People are beginning to build critical distributed applications,
particularly security applications, using system-level
provenance. However, it is difficult for them to know how to interpret
results or implement queries to detect activity on different
systems. Both potential users and developers of all three considered
provenance recording tools have participated in the design of
ProvMark.  To clarify when, and to whom, ProvMark is useful, in this
section we outline several common use cases.  In each case, ProvMark
automates a central, labor-intensive step: namely, running tools to
extract provenance, graphs and analyzing the graphs produced by a tool
to identify a target activity.  The first two use cases are real
(albeit lightly fictionalized) situations where we have used ProvMark.
The other two are more speculative, but illustrate how ProvMark could
be used for testing or exploration.

\paragraph{Tracking failed calls}

Alice, a security analyst, wants to know which provenance recorders
track syscalls that fail due to access control violations, since these
calls may be an indication of an attack or surveillance attempt.  She
can write small benchmark programs that capture various access control
failure scenarios; most only take a few minutes to write, by modifying
other, similar benchmarks for successful calls.  Once this is done, ProvMark can
run all of them to produce benchmark results.  For example, Alice
considers what happens if a non-privileged user unsuccessfully
attempts to overwrite \texttt{/etc/passwd} by renaming another file.

By default SPADE installs Linux Audit rules that only report on
successful system calls, so SPADE records no information in this case.
OPUS monitors system calls via intercepting C library calls, so it
knows whether a call is being attempted, and typically generates some
graph structure even for unsuccessful calls.  For example, the result
of a failed \syscall{rename} call has the same structure as shown in
Figure~\ref{fig:rename}, but with a different return value property of -1
instead of 0.  Finally, CamFlow can in principle monitor failed system
calls, particularly involving permission checks, but does not do so in
this case.  Alice concludes that for her application, OPUS may provide
the best solution, but resolves to discuss this issue with the SPADE
and CamFlow developers as well.

\paragraph{Configuration validation}

Bob, a system administrator, wants to make sure that an installation
of SPADE is configured correctly to match a security policy.  SPADE
(like the other systems) has several configuration parameters, to
allow greater or lesser levels of detail, coalescing similar
operations (such as repeated reads or writes), or enabling/disabling
versioning.  These also affect performance and system overhead, so Bob
wants to ensure that enough information is recorded to enable
successful audits, while minimizing system load.

Bob can use ProvMark to benchmark alternative configurations of
SPADE. For example, SPADE provides a flag \texttt{simplify} that is
enabled by default.  Disabling \texttt{simplify} causes
\syscall{setresgid} and \syscall{setresuid} (among others) to be
explicitly monitored, and Bob wants to ensure these calls are tracked.
However, on doing so, Bob also uncovered a minor bug: when
\syscall{simplify} is disabled, one of the properties of a background
edge is initialized to a random value, which shows up in the benchmark
as a disconnected subgraph.  The bug was reported to the SPADE
developers and quickly fixed.

SPADE also provides various \emph{filters} and \emph{transformers}
which perform pre-processing or post-processing of the stored
provenance respectively.  Bob also experimented with one of SPADE's filters,
\texttt{IORuns}, which controls whether runs of similar read or write
operations are coalesced into a single edge.  Surprisingly, Bob found that in the
benchmarked version of SPADE, enabling this filter had no effect.
This turned out to be due to an inconsistency in the property names
used by the filter vs. those generated by SPADE.  This has also now
been fixed.

\paragraph{Regression testing}

Charlie, a developer of provenance recording tool XYZTrace, wants to
be able to document the level of completeness of XYZTrace to
(potentially skeptical) users.  ProvMark can be used for regression
testing, by recording the graphs produced in a given benchmarking run,
and comparing them with the results of future runs, using the same
code for graph isomorphism testing ProvMark already uses during
benchmarking.  Charlie writes a simple script that stores the
benchmark graphs (as Datalog) from previous runs, and whenever the
system is changed, a new benchmarking run is performed and the results
compared with the previous ones.  When differences are detected, if the changes are expected then the new version of the graph replaces the old one; if the changes are unexpected, this is investigated as a potential bug.

\paragraph{Suspicious activity detection}

Dora, a security researcher, wants to identify patterns in provenance
graphs that are indicative of a potential attack on the system.  She
compiles a set of scripts that carry out different kinds of attacks,
and configures CamFlow on a virtual machine.  She is particularly
interested in detecting privilege escalation events where an attacker
is able to gain access to new resources by subverting a privileged
process.  She instruments the scripts to treat the privilege
escalation step as the ``target activity''.  Using ProvMark, Dora can
obtain example provenance graphs that include the target activity or
exclude it.  If the graphs are not too large (e.g. hundreds rather
than thousands of nodes), ProvMark's default behavior will also be
able to compute the differences between the graphs with and without
the target activity.

\subsection{Recording}

The \emph{recording} subsystem runs the
provenance capture tools on test
programs.
This subsystem first prepares a staging directory that provides a consistent
environment for test execution.  
The recording subsystem then starts the provenance capture tool with
appropriate settings, captures the provenance generated by the tool,
and stops the tool afterwards.

The recording subsystem is the only one to interact directly with the
target provenance capture tools.  For each tool, we implemented a
script that configures the tool to capture the provenance of a
specific process, rather than recording all contemporaneous system
events. Recording multiple runs of the same process using CamFlow
was challenging in earlier versions because CamFlow only serialized
nodes and edges once, when first seen.  The current version (0.4.5)
provides a workaround to this problem that re-serializes the needed
structures when they are referenced later.  We also modified its
configuration slightly to avoid tracking ProvMark's own behavior.  We
otherwise use the default configuration for each system; we refer to
these configurations as the \emph{baseline} configurations.  

The recording subsystem is used to record provenance graphs for the
foreground and background variants of each benchmark.  Provenance can
include transient data that varies across runs, such as timestamps, so
we typically record multiple runs of each program and filter out the
transient information in a later \emph{generalization} stage,
described below.

Some of the provenance collecting tools were originally designed to
record whole system provenance. The tools start recording when the
machine is started and only stops when the machine is shut
down. This behaviour ensures that provenance recording covers the
entire operating system session. As we need to obtain repeatable
results from several recording attempts for generalization and
benchmark generation, we need to restart the recording section
multiple times during the provenance collection. This may interfere with
the results of some tools as they are not originally designed for
frequent restarting of recording sessions. Thus the recording
subsystem aims to manage the collecting tools by inserting and
managing timeouts between the sessions. For example, we usually obtain
stable results from SPADE by inserting timeouts to wait for successful
completion of SPADE's graph generation process; even so, we sometimes
stop too early and obtain inconsistent results leading to mismatched
graphs.  Similarly, using CamFlow, we sometimes experience small
variations in the size or structure of the results for reasons we have
not been able to determine.  In both cases we deal with this by
running a larger number of trials and retaining the two smallest
consistent results (as discussed later).  For OPUS, any two runs are
usually consistent, but starting OPUS and loading data into or out of
Neo4j are time-consuming operations.

\subsection{Transformation}

Different systems output their provenance graphs in different formats.
For example, SPADE supports Graphviz DOT format and Neo4J storage
(among others), OPUS also supports Neo4J storage, and CamFlow supports
W3C PROV-JSON~\cite{provjson} as well as a number of other storage or
stream processing backends.  CamFlow also can be used instead of Linux
Audit as a reporter to SPADE, though we have not yet experimented with
this configuration.
To streamline the remaining stages, we translate these three
formats to a common representation.  Unlike the recording stage, this
stage really is straightforward, but we will describe the common
format in detail because it is important for understanding the
remaining stages.

\setlength{\abovecaptionskip}{0pt}
\begin{figure}[tb]
\lstset{caption={Datalog Graph
    Format},frame=single,label={potassco:index}}
\begin{lstlisting}
  Node n<gid>(<nodeID>,<label>)
  Edge e<gid>(<edgeID>,<srcID>,<tgtID>,<label>)
  Property p<gid>(<nodeID/edgeID>,<key>,<value>)
\end{lstlisting}

\end{figure}

The common target format is a logical representation of  property
graphs, in which nodes
and edges can have labels as well as associated properties (key-value
dictionaries).  Specifically, given a set $\Sigma$ of node and edge
labels, $\Gamma$ of property keys, and $D$ of data values, we consider
property graphs $G = (V,E,src,tgt,lab,prop)$ where $V$ is a set
of vertex identifiers and $E$ a set of edge identifiers; these are
disjoint ($V \cap E = \emptyset$). Further, $src,tgt: E \to V$ maps each edge $e$
to its source and target nodes, respectively, 
$lab : V \cup E \to \Sigma$ maps each node or edge $x$ to its
label $lab(x) \in \Sigma$, and
$prop : (V \cup E) \times \Gamma \to D$ is a partial function such
that for a given node or edge $x$, then $prop(x,p)$ (if defined) is
the value for property $p\in \Gamma$.  In practice, $\Sigma$, $\Gamma$
and $D$ are each sets of strings.

\setlength{\abovecaptionskip}{0pt}
\begin{figure}[tb]
  \begin{center}
    \includegraphics[scale=0.35]{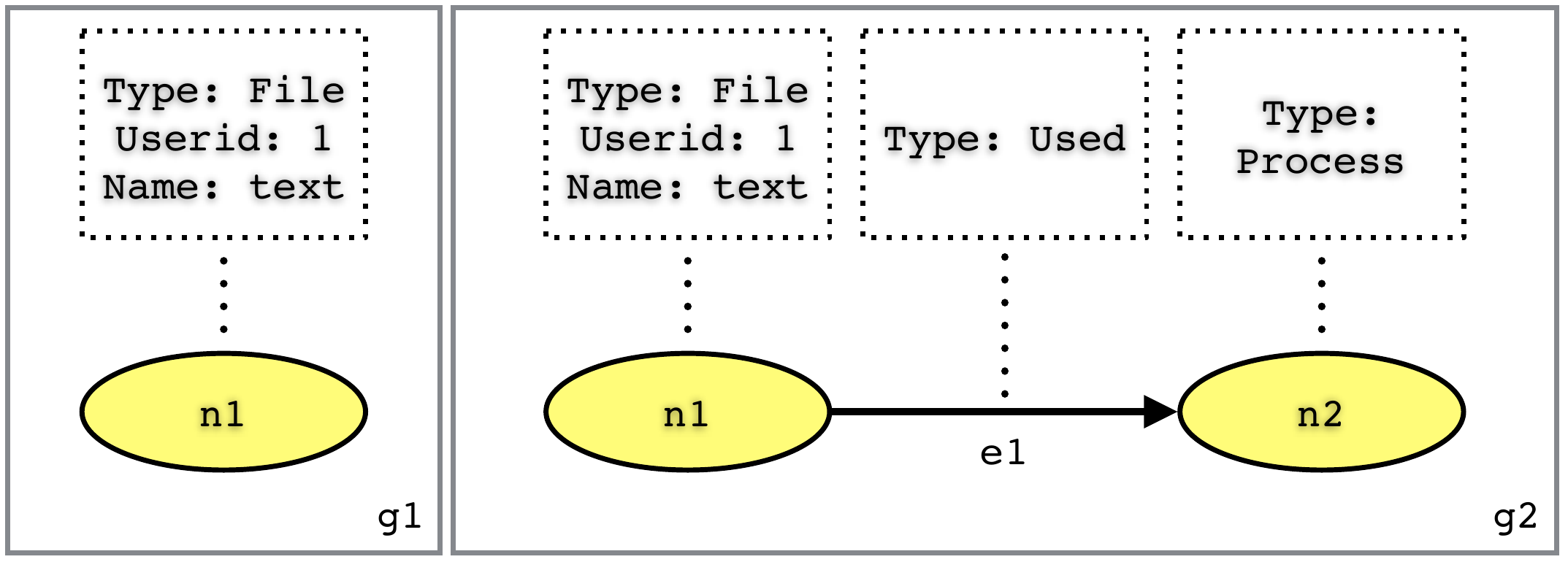}
    \captionof{figure}{Sample Graphs}\label{graph:sample}
  \end{center}

\lstset{caption={Datalog Format for Figure
    \ref{graph:sample}},frame=single,label={potassco:sample}}
\begin{lstlisting}
  ng1(n1,"File").
  pg1(n1,"Userid","1").
  pg1(n1,"Name","text").

  ng2(n1,"File").
  ng2(n2,"Process").
  pg2(n1,"Userid","1").
  eg2(e1,n1,n2,"Used").
  pg2(n1,"Name","text").
\end{lstlisting}
\end{figure}
\setlength{\abovecaptionskip}{10pt}

For provenance graphs, following the W3C PROV vocabulary, the node
labels are typically \emph{entity}, \emph{activity} and \emph{agent},
edge labels are typically relations such as \emph{wasGeneratedBy} and
\emph{used}, and properties are either PROV-specific property names or
domain-specific ones, and their values are typically strings.
However, our representation does not assume the labels and properties
are known in advance; it works with those produced by the tested system.

We represent property graphs as
sets of logical facts using a Prolog-like syntax called
\emph{Datalog}~\cite{ahv}, which is often used to represent relational data in
logic programming (as well as databases~\cite{ahv} and networking~\cite{green13ftdb}).

The generic form of the Datalog graph format we use is shown as
Listing \ref{potassco:index}. We assume a fixed string \verb|gid| used
to uniquely identify a given graph as part of its node, edge and label
relation names. Each node $v \in V$ is represented as a fact
$n_{gid}(v, lab(v))$.  Likewise, each edge $e = (v,w) \in E$ is represented as a
fact $e_{gid}(e, src(e), tgt(e), lab(e))$.  Finally, if a node or edge $x$ has property $p$
with value $s$, we represent this with the fact
$p_{gid}(x, p, s)$.  Two sample graphs are shown in Figure \ref{graph:sample}
and their Datalog representations in Listing
\ref{potassco:sample}.

All the remaining stages, including the final result, work on the
Datalog graph representation, so these stages are independent of
the provenance capture tool and its output format.  The Datalog
representation can easily be visualized.

\subsection{Graph Generalization}

The third subsystem performs \emph{graph
  generalization}.  Recall that the recording stage produces several
graphs for a given test program.  We wish to identify a single,
representative and general graph for each test program.  To formalize
these notions, we adapt the notion of \emph{graph isomorphism} to
property graphs:
$G_1$ is isomorphic to $G_2$ if there is an invertible function $h :
V_1 \cup E_1 \to V_2\cup E_2$
such that 
\begin{enumerate}
\item $h(src_1(e)) = src_2(h(e))$ and $h(tgt_1(e)) = tgt_2(h(e))$,
  i.e. $h$ preserves edge relationships
\item $lab_1(x) = lab_2(h(x))$, i.e. $h$ preserves node and edge labels
\item $prop_1(x,k,v) = prop_2(h(x),k,v)$, i.e. $h$ preserves
  properties.
\end{enumerate}
 Moreover, we say that $G_1$ and $G_2$ are \emph{similar}
  if only the first two conditions hold, that is, if $G_1$ and $G_2$
  have the same shape but possibly different properties.

We assume that over sufficiently many recording trials, there will be at least
two \emph{representative} ones, which are similar to each other.
Identifying two representative runs is complicated by the fact that
there might be multiple pairs of similar graphs.  We discuss a
strategy for identifying an appropriate pair below.
To obtain two representative graphs, we first consider all of the
trial runs and partition them into similarity classes.  We first
discard all graphs that are only similar to themselves, and consider
these to be failed runs.  Among the remaining similarity classes, we
choose a pair of graphs whose size is smallest.  Picking the two
largest graphs also seems to work; the choice seems arbitrary.  However,
picking the largest background graph and the smallest foreground graph
leads to failure if the extra background structure is not found in the
foreground, while making the opposite choice leads to extra structure
being found in the difference.

Given two similar graphs, the generalization stage identifies the
property values that are consistent across the two graphs and removes
the transient ones.
The generalization stage searches for a matching of nodes and edges of the
two graphs that minimizes the number of different properties.  We 
assume that the differences are transient data and discard them.  We
would like to minimize the number of differences, that is, match as
many properties as possible.

Similarity and generalization are instances of the \emph{graph
  isomorphism} problem, whose status (in P or NP-complete) is
unknown~\cite{arvind05beatcs}.  We solve these problems using an
Answer Set Programming (ASP)~\cite{gebser11aicom} specification that
defines the desired matching between the two graphs.  ASP is a
decidable form of logic programming that combines a high-level logical
specification language with efficient search and optimization
techniques analogous to SAT solving or integer linear programming.  An answer
set program specifies a set of possible \emph{models}, which
correspond to solutions to a problem.  The specification is a
collection of logical formulas that define when a model is a possible
solution. In our
setting, the models consist of matching relations between two graphs,
satisfying the requirements of graph similarity.  ASP is well-suited to this problem
because the graphs and specification can be represented concisely,
and (as we shall see) the ASP solver can efficiently find solutions.

The problem specification is a logic program defining when a
binary relation forms an isomorphism between the graphs.  
The code in Listing~\ref{fig:gi} defines graph isomorphism in ASP.
ASP specifications consist of rules that constrain the possible
solution set.  The first four lines specify that $h$ can relate any
node in $G_1$ to any node in $G_2$ and vice versa, and similarly for
edges.  The next two lines constrain $h$ to be a 1-1 function.  (Rules of
the form \verb|:- A,B,C.| can be read as ``It cannot be the case that
\verb|A|, \verb|B|, and \verb|C| hold''.)  The remaining three pairs
of lines specify that $h$ must preserve node and edge labels and the
source and targets of edges.

\begin{figure}[t]
\lstset{frame=single,caption={Graph similarity},label={fig:gi}}
\begin{lstlisting}
{h(X,Y) : n2(Y,_)} = 1 :- n1(X,_).
{h(X,Y) : n1(X,_)} = 1 :- n2(Y,_). 
{h(X,Y) : e2(Y,_,_,_)} = 1 :- e1(X,_,_,_).
{h(X,Y) : e1(X,_,_,_)} = 1 :- e2(Y,_,_,_).

:- X <> Y, h(X,Z), h(Y,Z).
:- X <> Y, h(Z,Y), h(Z,X).

:- n1(X,L), h(X,Y), not n2(Y,L).
:- n2(Y,L), h(X,Y), not n1(X,L).

:- e1(E1,_,_,L), h(E1,E2), not e2(E2,_,_,L).
:- e2(E2,_,_,L), h(E1,E2), not e1(E1,_,_,L).

:- e1(E1,X,_,_), h(E1,E2), e2(E2,Y,_,_), not h(X,Y).  
:- e1(E1,_,X,_), h(E1,E2), e2(E2,_,Y,_), not h(X,Y).
\end{lstlisting}
\end{figure}

This specification can be solved using \texttt{clingo}, an efficient
ASP solver~\cite{gebser11aicom}.  As ASP is a kind of logic
programming, it helps ProvMark to determine an optimal matching
solution among two graphs by searching for a model that satisfies the
specification. We use the resulting matching to determine which
properties are common across both graphs and discard the others.  We
perform generalization independently on the foreground and background
graphs.  The outputs of this stage are the two generalized graphs
representing the invariant activity of the foreground and background
programs respectively.

\subsection{Graph Comparison}

The fourth and last subsystem is \emph{graph comparison}. Its purpose
is to match the background graph to a subgraph of the foreground
graph; the unmatched part of the foreground graph corresponds to the
target activity.

Because provenance recording is generally monotonic (append-only), we expect that the generalized provenance graph for the background
program is a subgraph of the generalized provenance graph for the
foreground program, so there should be a one-to-one matching from the
nodes and edges in the background graph to the foreground graph.  This
graph matching problem is a variant of the subgraph isomorphism
problem, a classical NP-complete problem~\cite{cook71stoc}.
We again solve these problems using ASP, taking advantage of the fact
that ASP solvers can search for optimal solutions according to some
cost model.

Given two graphs, $G_1$ and
$G_2$, the solver finds a matching from the nodes and edges of $G_1$
to those of $G_2$ that identifies a subgraph of $G_2$ isomorphic to
$G_1$ and minimizes the number of mismatched properties. 
Listing~\ref{fig:sg-approx} shows the approximate subgraph
optimization problem specification.  It is related to graph
similarity, but only requires nodes/edges in $G_1$ to be matched
to one in $G_2$ and not the reverse.  Additionally, the last four
lines define the cost of a matching as the number of properties
present in $G_1$ with no matching property in $G_2$, and this cost is
to be
minimized.

\begin{figure}[tb]
  \lstset{frame=single,caption={Approximate subgraph
      isomorphism},label={fig:sg-approx}}
  \begin{lstlisting}
{h(X,Y) : n2(Y,_)} = 1 :- n1(X,_).  
{h(X,Y) : e2(Y,_,_,_)} = 1 :- e1(X,_,_,_).

:- X <> Y, h(X,Z), h(Y,Z).  
:- X <> Y, h(Z,Y), h(Z,X).

:- n1(X,L), h(X,Y), not n2(Y,L).  
:- e1(E1,_,_,L), h(E1,E2), not e2(E2,_,_,L).

:- e1(E1,X,_,_), h(E1,E2), e2(E2,Y,_,_), not h(X,Y).  
:- e1(E1,_,X,_), h(E1,E2), e2(E2,_,Y,_), not h(X,Y).

cost(X,K,0) :- p1(X,K,V), h(X,Y), p2(Y,K,V).
cost(X,K,1) :- p1(X,K,V), h(X,Y), p2(Y,K,W), V <> W.
cost(X,K,1) :- p1(X,K,V), h(X,Y), not p2(Y,K,_).

#minimize { PC,X,K : cost(X,K,PC) }.
  \end{lstlisting}
\end{figure}

We again use
the \texttt{clingo} solver to find an optimal matching. Once we have
found such a matching, the graph comparison subsystem subtracts the
matched nodes and edges in the background activity from the foreground
graph.  This is essentially a set difference operation between the
nodes and edges of the graphs, but we also retain any nodes that are
sources or targets of edges in the difference; these are visualized as
green or gray \emph{dummy nodes}.

\section{Demonstration}
\label{sec:result}
We present a demonstration of using ProvMark to investigate and
compare the behavior of the different capture systems. These results
pertain to SPADEv2 (tag \emph{tc-e3}) and OPUS version 0.1.0.26
running on Ubuntu 16.04 and CamFlow version 0.4.5 running on Fedora
27.

Unix-like operating systems support over three hundred system calls.
We focus on a subset of 22 common system call families shown in
Table~\ref{table:syscall}, including the most common file and process
management calls.  We prepared benchmarking programs for these
syscalls, along with tests for each one to ensure that the target
behavior was performed successfully.  Note that the same system call
may display different behavior using different parameters and system
states; we focus on common-case behavior here, but handling other
scenarios such as failure cases is straightforward as outlined in
Section~\ref{sec:usecases}.

\begin{table}[tb]
  \center
  \begin{tabular}{|l|l|p{5.5cm}|}
\hline
1 &Files & \syscall{close}, \syscall{creat}, \syscall{dup[2,3]},  \syscall{[sym]link[at]}, \syscall{mknod[at]}, \syscall{open[at]}, 
        \syscall{[p]read}, \syscall{rename[at]}, \syscall{[f]truncate},
        \syscall{unlink[at]}, \syscall{[p]write}
\\ \hline
2& Processes & \syscall{clone}, \syscall{execve}, \syscall{exit}, \syscall{[v]fork}, \syscall{kill}\\\hline
3& Permissions & \syscall{[f]chmod[at]}, \syscall{[f]chown[at]}, \syscall{set[re[s]]gid}, \syscall{set[re[s]]uid}\\\hline
4& Pipes & \syscall{pipe[2]}, \syscall{tee}\\\hline
\end{tabular}
  \caption{Benchmarked syscalls}\label{table:syscall}
  \normalsize
\end{table}

We cannot show all of the benchmark graphs, nor can we show the graphs
at a readable size; the full results are available for inspection
online~\cite{provmark2018}.  Table \ref{tab:validation} summarizes the
results.  In this table, ``ok'' means the graph is correct (according
to the system's developers), and ``empty'' means the foreground and
background graphs were similar and so the target activity was undetected.  The table includes notes indicating
the reason for emptiness.  Also, Table~\ref{table:results} shows
selected graphs (again, images are links to scalable online images
with full details) for the benchmark result. Although the property
keys and values are omitted, we can compare the graph structure and
see that the three tools generate very different structures for the
same system calls. This demonstrates one of the important motivations
for expressiveness benchmarking. In these graph results, the yellow
ovals represent artifacts which includes files, pipes or other
resources. The blue rectangles represent processes involved in the
named system calls. The green or gray ovals represent dummy nodes
which stand for pre-existing parts of the graph. We retain these
dummy nodes to make the result a complete graph.

In this section we present and discuss the benchmarking outputs for
several representative syscalls, highlighting minor bugs found and
cases where the behavior of individual tools or the variation among
different tools was surprising.

\begin{table}[tb]
  \centering
  \begin{tabular}{|c|c|c|c|c|}
\hline
Group& syscall & SPADE & OPUS & CamFlow\\\hline
1& close & ok & ok & empty (LP)\\
1& creat & ok & ok & ok\\
1&  dup & empty (SC) & ok & empty (NR) \\
1& dup2 & empty (SC) & ok & empty (NR) \\
1& dup3 & empty (SC) & ok & empty (NR) \\
1& link & ok & ok & ok\\
1& linkat & ok & ok & ok\\
1& symlink & ok & ok & empty (NR)\\
1& symlinkat & ok & ok & empty (NR) \\
1& mknod & empty (NR) & ok & empty (NR)\\
1& mknodat & empty (NR) & empty (NR) & empty (NR)\\
1& open & ok & ok & ok\\
1& openat & ok & ok & ok\\
1& read & ok & empty  (NR) & ok\\
1& pread & ok & empty (NR) & ok\\
1& rename & ok & ok & ok\\
1& renameat & ok & ok & ok\\
1& truncate & ok & ok & ok\\
1& ftruncate & ok & ok & ok\\
1&unlink & ok & ok & ok\\
1&unlinkat & ok & ok & ok\\
1&write & ok & empty (NR) & ok\\
1& pwrite & ok & empty  (NR) & ok\\
\hline
2&clone & ok & empty (NR) & ok\\
2& execve & ok & ok & ok\\
2& exit & empty (LP)  &empty (LP)  & empty (LP)\\
2& fork & ok & ok & ok\\
2& kill & empty (LP)  &empty (LP)  & empty (LP)\\ 
2& vfork & ok (DV) & ok & ok\\
\hline
3&chmod & ok & ok & ok\\
3&fchmod & ok & empty (NR) & ok\\
3&fchmodat & ok & ok & ok\\
3&chown & empty (NR) & ok & ok\\
3&fchown & empty (NR) & empty (NR) & ok\\
3&fchownat & empty (NR)& ok & ok\\
3&setgid & ok & ok & ok\\
3&setregid & ok & ok & ok\\
3&setresgid & empty (SC) & empty (NR) & ok\\
3&setuid & ok & ok & ok\\
3&setreuid & ok & ok & ok\\
 3&setresuid & ok (SC) & empty (NR) & ok\\
\hline
4&pipe & empty (NR) & ok & empty (NR)\\
4&pipe2 & empty (NR) & ok & empty (NR)\\
4&tee & empty (NR) & empty (NR) & ok\\
\hline
\end{tabular}
\\\smallskip
\begin{tabular}{|c|p{7cm}|}
\hline  Note& Meaning\\\hline
NR & Behavior not recorded (by default configuration)\\
SC & Only state changes monitored\\
LP & Limitation in ProvMark\\
DV & Disconnected \syscall{vfork}ed process\\\hline
\end{tabular}
  \caption{Summary of validation results}
  \label{tab:validation}
\end{table}
\normalsize

\subsection{File system}

File system calls such as \syscall{creat}, \syscall{open},
\syscall{close} are generally covered well by all of the tools, but
they each take somewhat different approaches to representing even
simple behavior.  For example, for the \syscall{open} call, SPADE adds
a single node and edge for the new file, while CamFlow creates a node
for the file object, a node for its path, and several edges linking
them to each other and the opening process, and OPUS creates four new
nodes including two nodes corresponding to the file.  On the other
hand, reads and writes appear similar in both SPADE and CamFlow, while
by default, OPUS does not record file reads or
writes. For \syscall{close}, CamFlow
records the underlying kernel data structures eventually being freed,
but ProvMark does not reliably observe this.

The \syscall{dup[2,3]} syscall duplicates a file descriptor so that
the process can use two different file descriptors to access the same
open file.  SPADE and CamFlow do not record this call directly, but the
changes to the process's file descriptor state can affect future file
system calls that are monitored.  OPUS does record \syscall{dup[2,3]}.
The two added nodes are not directly connected to
each other, but are connected to the same process node in the
underlying graph.  One component records the system call itself and
the other one records the creation of a new resource (the duplicated
file descriptor).  

The \syscall{link[at]} call is recorded by all three systems but
\syscall{symlink[at]} is not recorded by CamFlow 0.4.5.  Likewise,
\syscall{mknod} is only recorded by OPUS, and \syscall{mknodat} is not
recorded by any of the systems.

The \syscall{rename[at]} call, as shown in the introduction, illustrates
how the three systems record different graph structure for this
operation.  SPADE represents a rename using two nodes for the new and
old filenames, with edges linking them to each other and to the
process that performed the rename.  OPUS creates around a dozen nodes,
including nodes corresponding to the rename call itself and the new and
old filename.  CamFlow represents a rename as adding a new path
associated with the file object; the old path does not appear in the
benchmark result.

\setlength{\abovecaptionskip}{0pt}
\begin{table*}[tb]
\caption{Example benchmark results for SPADE, OPUS and CamFlow.
  Images are links to scalable online images with property labels.}\label{table:results}
\begin{center}
\setlength\tabcolsep{1.5pt}
\begin{tabular}{c|c|c|c|c|c|c}

& open & read & write & dup & setuid & setresuid \\ \hline

SPADE
& 
	\href{https://provmark2018.github.io/sampleResult/spade/open.svg}{
    		\includegraphics[align=c,scale=0.20]{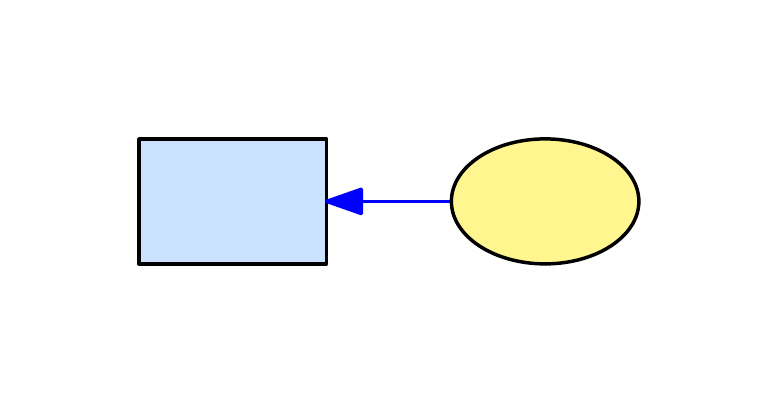}
    	}
&
	\href{https://provmark2018.github.io/sampleResult/spade/read.svg}{
    		\includegraphics[align=c,scale=0.20]{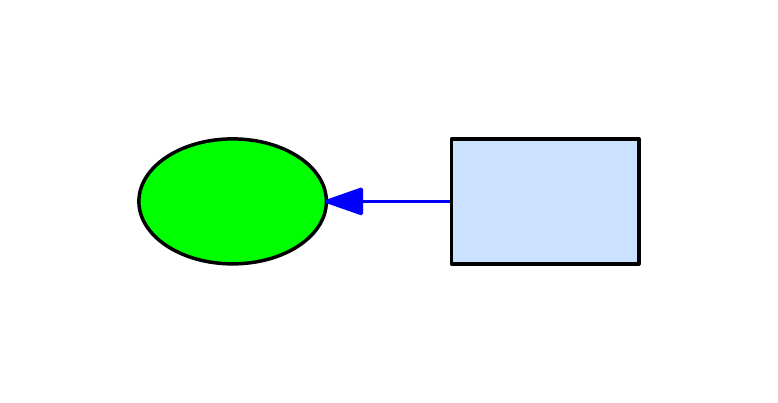}
    	}
&
	\href{https://provmark2018.github.io/sampleResult/spade/write.svg}{
    		\includegraphics[align=c,scale=0.20]{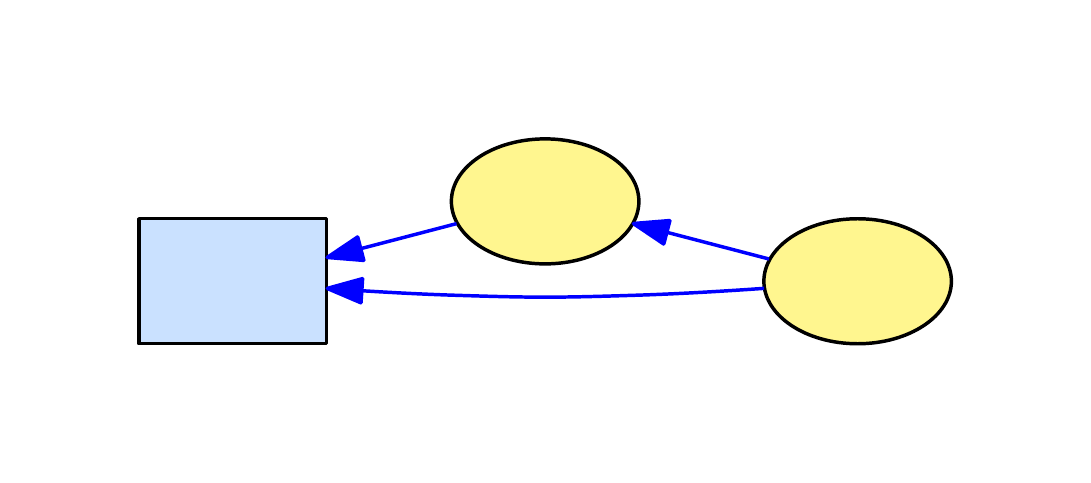}
    	}
&
	Empty
&
	\href{https://provmark2018.github.io/sampleResult/spade/setuid.svg}{
    		\includegraphics[align=c,scale=0.20]{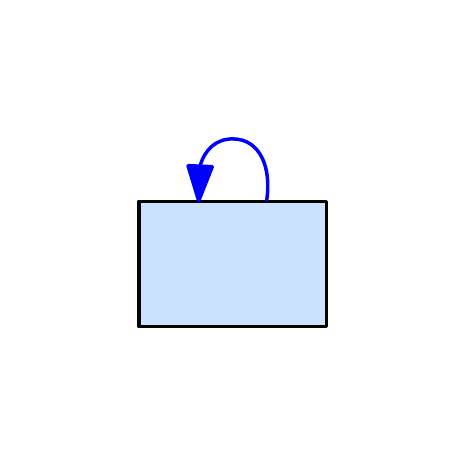}
    	}
&
	\href{https://provmark2018.github.io/sampleResult/spade/setresuid.svg}{
    		\includegraphics[align=c,scale=0.20]{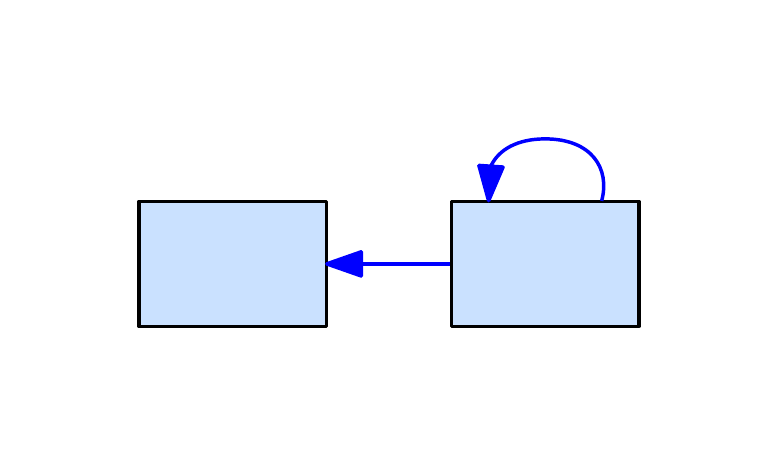}
    	}
\\ \hline

OPUS
& 
	\href{https://provmark2018.github.io/sampleResult/opus/open.svg}{
    		\includegraphics[align=c,scale=0.20]{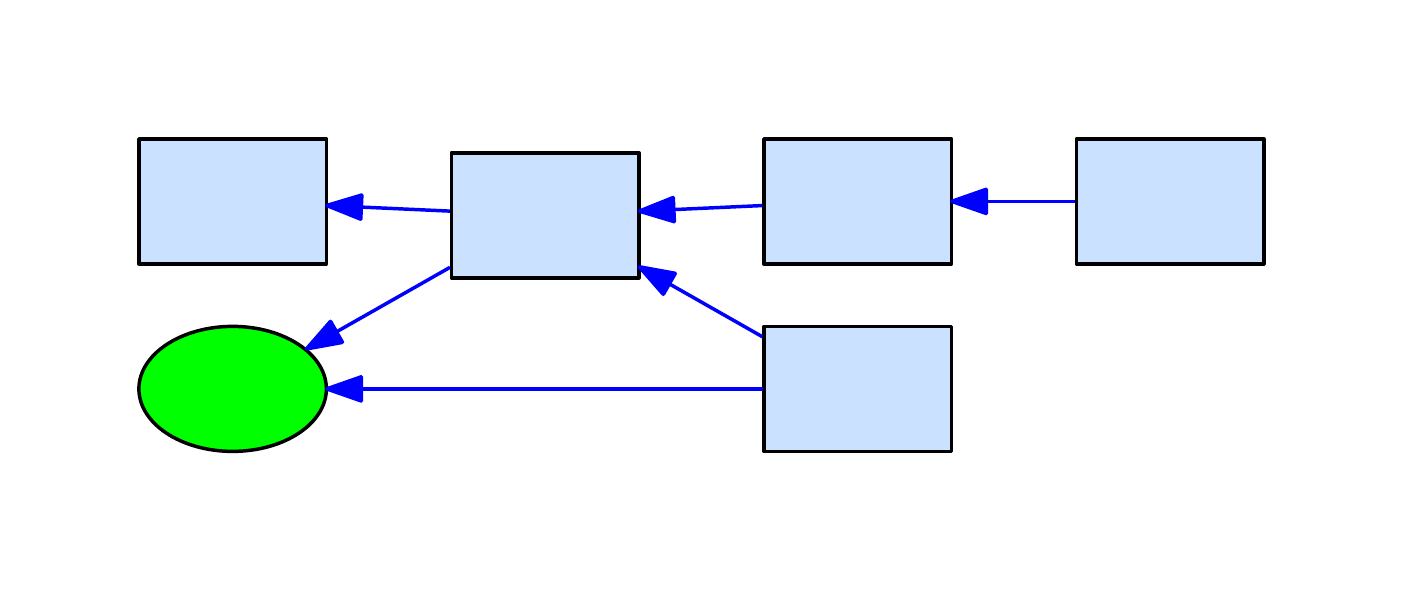}
    	}
&
	Empty
&
	Empty
&
	\href{https://provmark2018.github.io/sampleResult/opus/dup.svg}{
    		\includegraphics[align=c,scale=0.20]{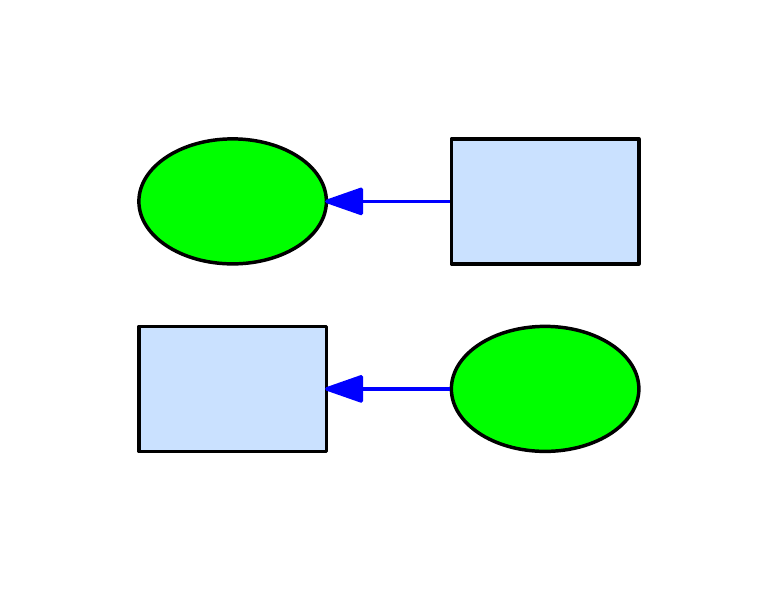}
    	}
&
	\href{https://provmark2018.github.io/sampleResult/opus/setuid.svg}{
    		\includegraphics[align=c,scale=0.20]{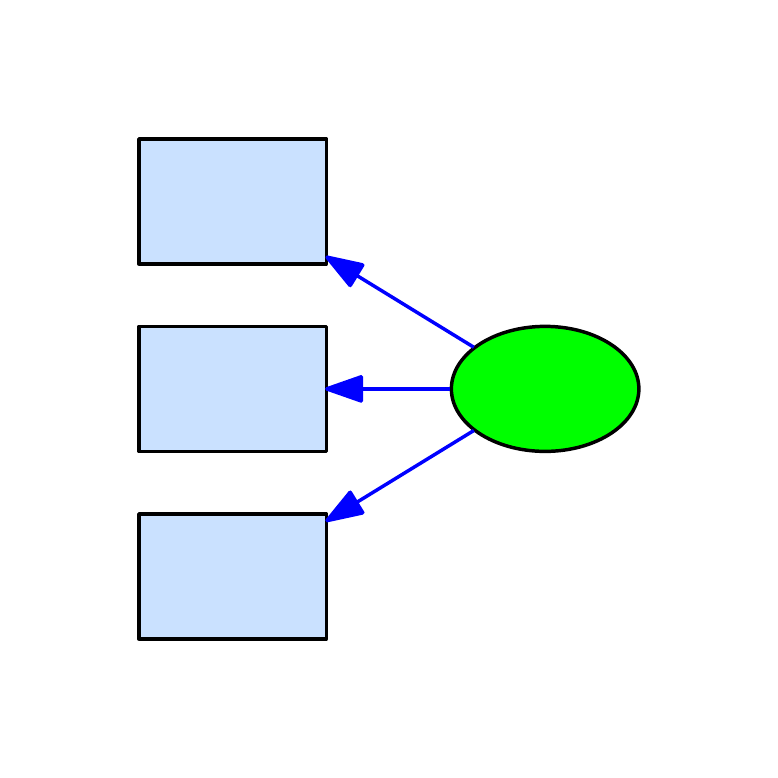}
    	}
&
	Empty
\\ \hline

CamFlow
& 
	\href{https://provmark2018.github.io/sampleResult/camflow/open.svg}{
    		\includegraphics[align=c,scale=0.20]{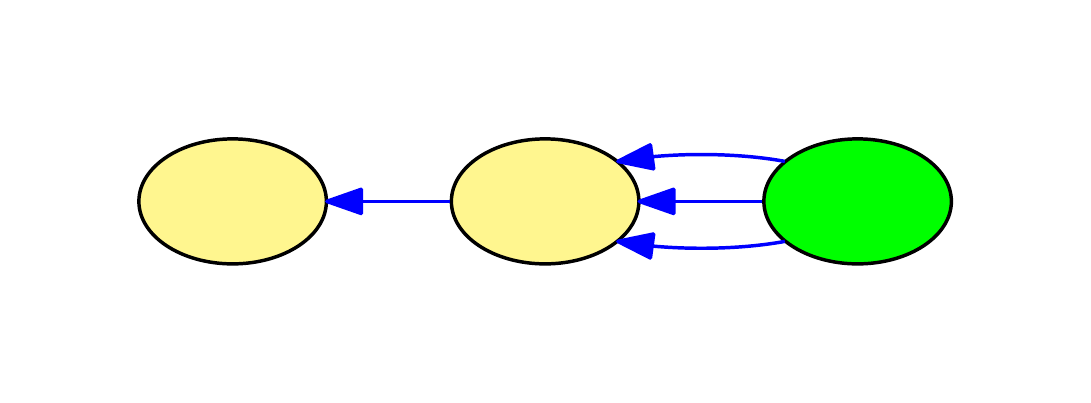}
    	}
&
	\href{https://provmark2018.github.io/sampleResult/camflow/read.svg}{
    		\includegraphics[align=c,scale=0.20]{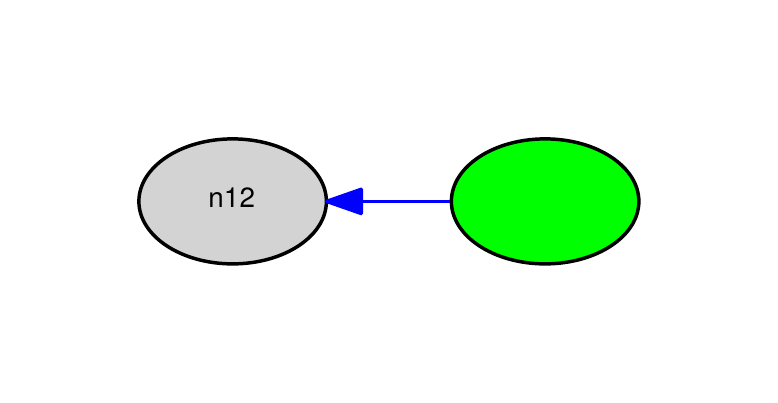}
    	}
&
	\href{https://provmark2018.github.io/sampleResult/camflow/write.svg}{
    		\includegraphics[align=c,scale=0.20]{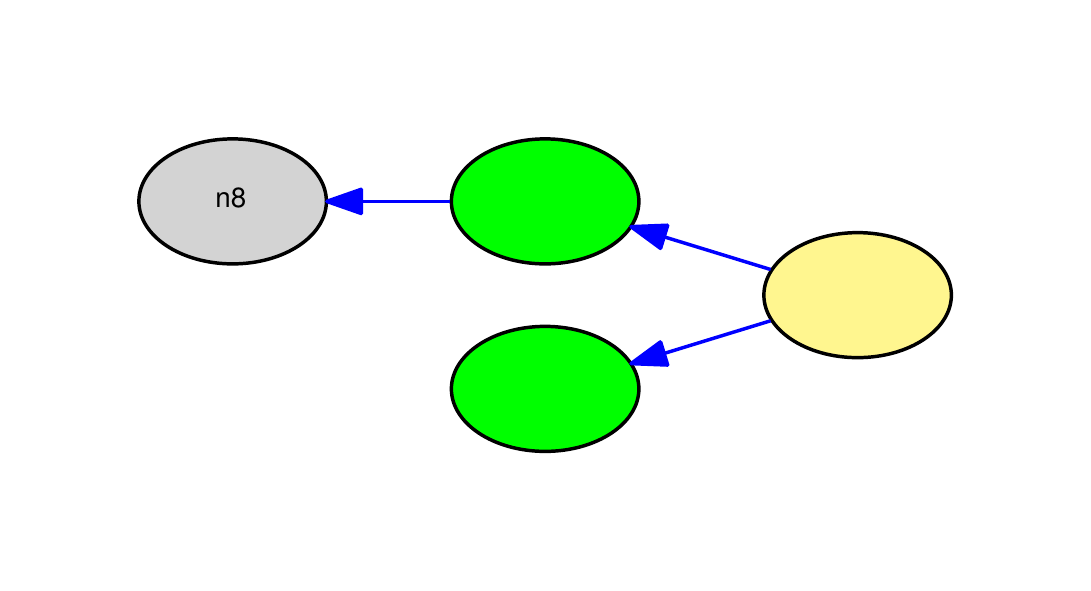}
    	}
&
	Empty
&
	\href{https://provmark2018.github.io/sampleResult/camflow/setuid.svg}{
    		\includegraphics[align=c,scale=0.20]{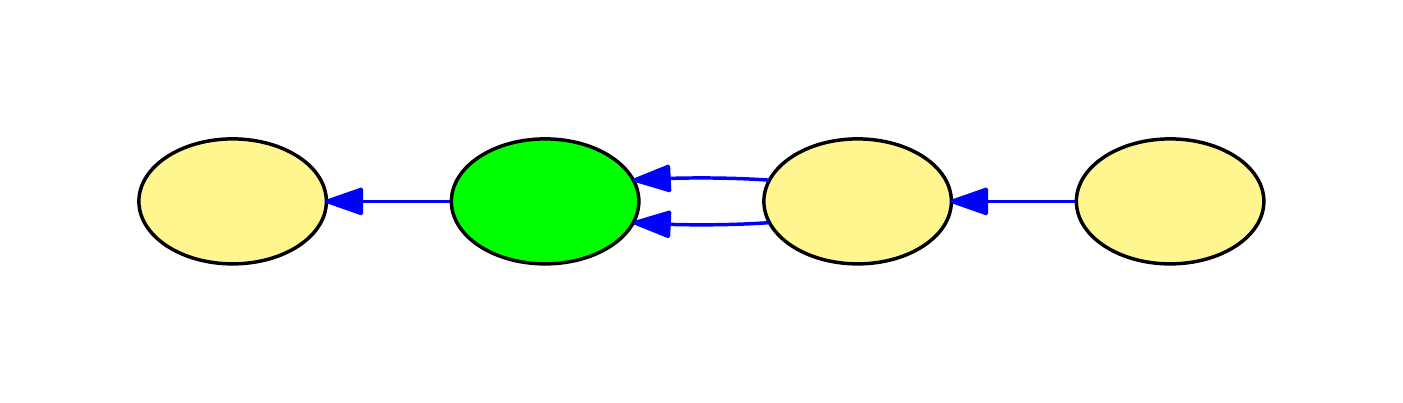}
    	}
&
	\href{https://provmark2018.github.io/sampleResult/camflow/setresuid.svg}{
    		\includegraphics[align=c,scale=0.20]{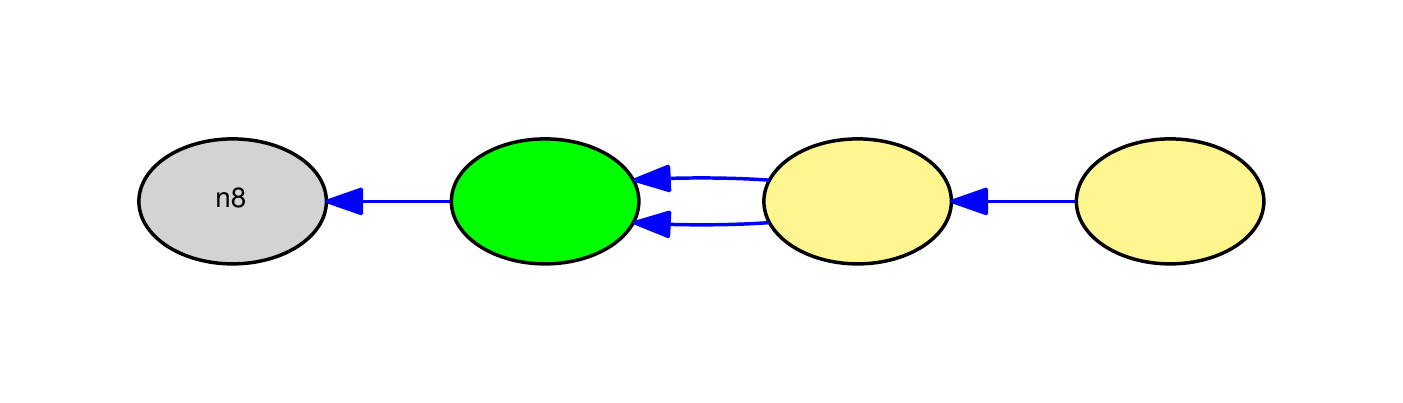}
    	}
  \end{tabular}
  \setlength\tabcolsep{6pt}
\end{center}
\end{table*}
\setlength{\abovecaptionskip}{10pt}

\subsection{Process management}

The process management operations start processes (\syscall{[v]fork},
\syscall{clone}), replace a process's memory space with a new
executable and execute it (\syscall{execve}), or terminate a process
(\syscall{kill}, \syscall{exit}). 

Process creation and initialization calls are generally monitored by
all three tools,
except that OPUS does not appear to monitor \syscall{clone}.
Inspecting the results manually shows significant differences,
however.  For example, the SPADE benchmark graph for \syscall{execve} is
large, but has just a few nodes for OPUS and CamFlow.  On
the other hand, the \syscall{fork} and \syscall{vfork} graphs are
small for SPADE and CamFlow and large for OPUS.  This may indicate the
different perspectives of these tools: SPADE relies on the activity
reported by Linux Audit, while OPUS relies on intercepting C library
calls outside the kernel.

Another interesting observation is that the SPADE benchmark results
for \syscall{fork} and \syscall{vfork} differ.  Specifically, for
\syscall{vfork}, SPADE represents the forked process as a disconnected
activity node, i.e. there is no graph structure connecting the parent
and child process.  This is because Linux Audit reports system calls
on exit, while the parent process that called \syscall{vfork} is
suspended until the child process exits.  SPADE sees the
\syscall{vforked} child process in the log executing its own syscalls
before it actually sees the \syscall{vfork} that created the child
process.  

The \syscall{exit} and \syscall{kill} calls are not detected 
because they deviate from the assumptions our approach is based on.
A process always has an implicit \syscall{exit} at the end, while
killing a process means that it does not terminate normally.  Thus,
the \syscall{exit} and \syscall{kill} benchmarks are all empty.  We
plan to consider a more sophisticated approach that can benchmark these
calls in future work.

\subsection{Permissions}
We included syscalls that change file permissions
(such as \syscall{[f]chown[at]}, \syscall{[f]chmod[at]}) or process
permissions in this group.  According to its documentation, SPADE
currently records \syscall{[f]chmod[at]} but not
\syscall{[f]chown[at]}.  OPUS does not monitor \syscall{fchmod} or
\syscall{fchown} because from its perspective these calls only perform
read/write activity and do not affect the process's file descriptor
state, so as for other read/write activity OPUS does not record
anything in its default configuration.  CamFlow records all of these
operations.

In its default configuration, SPADE does not explicitly  record
\syscall{setresuid} or \syscall{setresgid}, but it does monitor
changes to these process attributes and records observed changes.  Our
benchmark result for \syscall{setresuid} is nonempty, reflecting an
actual change of user id, while our benchmark for \syscall{setresgid}
just sets the group id attribute to its current value, and so this activity
is not noticed by SPADE.  OPUS also does not track these two calls,
while CamFlow again tracks all of them.

\subsection{Pipes}
Pipes provide efficient interprocess communication.  The
\syscall{pipe[2,3]} call creates a pipe, which can be read or written
using standard file system calls, and \syscall{tee} duplicates
information from one pipe into another without consuming it.
Of the three systems, only OPUS records \syscall{pipe[2,3]} calls, while
only CamFlow currently records \syscall{tee}.

\section{System Evaluation}
\label{sec:eval}

In this section, we will evaluate ProvMark. We claim that the ProvMark
system makes it easy to test and compare the behavior of provenance
tracking systems and to assess their completeness and correctness.
The previous section gave some evidence that the results are useful
for understanding the different systems and what information they do
or do not collect.  In this section, we will evaluate the performance
and extensibility of ProvMark itself. For a fair comparison, all the
experiments are done in a virtual machine with 1 virtual CPU and 4GB
of virtual memory. Dependencies for all three provenance collecting
tools, including Neo4J and some needed Python libraries, are also
installed.  All virtual machines were hosted on a Intel i5-4570 3.2GHz
with 8GB RAM running Scientific Linux 7.5.

\subsection{System Performance}

We first discuss the performance of the ProvMark system. Some
performance overhead is acceptable for a benchmarking or testing tool;
however, we need to establish that ProvMark's strategy for solving
NP-complete subgraph isomorphism problems is effective in practice
(i.e. takes minutes rather than days).
In this section, we will show that the extra work done by ProvMark to
generate benchmarks from the original provenance result is acceptable.

We first report measurements of the recording time.  Since the number
of trials varies, we report the average time needed per trial. For
SPADE, recording took approximately 20s per trial.  For OPUS, the
recording time was around 28s per trial, and for CamFlow each trial
took around 10s.  We wish to emphasize that the recording time results
are \emph{not} representative of the recording times of these systems
in normal operation --- they include repeatedly starting, stopping, and
waiting for output to be flushed, and the waiting times are
intentionally conservative to avoid garbled results.  No conclusions
about the relative performance of the recording tools when used as intended should be
drawn.

In Figures~\ref{chart:timedspade}--\ref{chart:timedcamflow}, we
summarize the time needed for ProvMark to run five representative
syscalls using SPADE, OPUS, and CamFlow respectively.  The
bars are divided into three portions, representing the time needed for
the
transformation, generalization and comparison subsystems. Note that
the x-axes are not all to the same scale: in particular the
transformation, generalization and comparison times for OPUS are much
higher than for the other tools because of database startup and access
time, and because the graphs extracted from the database are larger.
Modifying OPUS to circumvent the database and serialize provenance
directly would avoid this cost, but we have chosen to use the tools as
they are to the greatest extent possible.

\begin{figure}[tb]
\begin{tikzpicture}
\begin{axis}[
    xbar stacked,
    xlabel={Time (seconds)},
    symbolic y coords={open,execve,fork,setuid,rename},
    xmin = 0,
    xmax = 3,
    y=-0.32cm,
	bar width=0.2cm,
	ytick=data,
    legend style={nodes={scale=0.75, transform shape}, legend columns=1}
    ]
\addplot+[xbar] plot coordinates {
		(.208,open)
		(.261,execve)
		(.233,fork)
		(.204,setuid)
                (.207,rename)
};
\addplot+[xbar] plot coordinates {
		(.13,open)
		(1.133,execve)
		(0.135,fork)
		(0.128,setuid)
                (.136,rename)
};
\addplot+[xbar] plot coordinates {
		(.043,open)
		(.109,execve)
		(.041,fork)
		(.044,setuid)
                (.044,rename)
};
\legend{Transformation,Generalization,Comparison}
\end{axis}
\end{tikzpicture}
\caption{Timing results: SPADE+Graphviz}\label{chart:timedspade}
\bigskip
\begin{tikzpicture}
\begin{axis}[
    xbar stacked,
    xlabel={Time (seconds)},
    symbolic y coords={open,execve,fork,setuid,rename},
    xmin = 0,
    xmax = 2000,
    y=-0.32cm,
	bar width=0.2cm,
	ytick=data,
    legend style={nodes={scale=0.75, transform shape}, legend columns=1}
    ]
\addplot+[xbar] plot coordinates {
		(364.132,open)
		(356.858,execve)
		(372.7,fork)
		(377.446,setuid)
                (355.253,rename)
};
\addplot+[xbar] plot coordinates {
		(18.318,open)
		(16.306,execve)
		(48.461,fork)
		(17.271,setuid)
                (21.461,rename)
};
\addplot+[xbar] plot coordinates {
		(2.221,open)
		(2.051,execve)
		(731.556,fork)
		(2.198,setuid)
                (2.72,rename)
};
\legend{Transformation,Generalization,Comparison}
\end{axis}
\end{tikzpicture}
\caption{Timing results: OPUS+Neo4J}\label{chart:timedopus}
\bigskip
\begin{tikzpicture}
\begin{axis}[
    xbar stacked,
    xlabel={Time (seconds)},
    symbolic y coords={open,execve,fork,setuid,rename},
    xmin = 0,
    xmax =3,
    y=-0.32cm,
	bar width=0.2cm,
	ytick=data,
    legend style={nodes={scale=0.75, transform shape}, legend columns=1}
    ]
\addplot+[xbar] plot coordinates {
		(.878,open)
		(.867,execve)
		(.941,fork)
		(.837,setuid)
                (.879,rename)
};
\addplot+[xbar] plot coordinates {
		(.218,open)
		(.226,execve)
		(.282,fork)
		(.219,setuid)
                (.191,rename)
};
\addplot+[xbar] plot coordinates {
		(.119,open)
		(.114,execve)
		(.218,fork)
		(.104,setuid)
                (.103,rename)
};
\legend{Transformation,Generalization,Comparison}
\end{axis}
\end{tikzpicture}
\caption{Timing results: CamFlow+ProvJson}\label{chart:timedcamflow}
\end{figure}

From the data in
Figures~\ref{chart:timedspade}--\ref{chart:timedcamflow} we can see
that the transformation stage is typically the most time-consuming part.
The transformation stage maps the different provenance output formats
to Datalog format.  The transformation time for OPUS is much longer
than for the other two systems.  This appears to be because OPUS
produces larger graphs (including recording environment variables),
and extracting them involves running Neo4j queries, which also has a
one-time JVM warmup and database initialization cost.

The time required for the generalization stage depends mainly on the
number of elements (nodes and edges) in the graph, since this stage
matches elements in pairs of graphs to find an isomorphic graph
matching. As we can see from the results, the generalization of
OPUS graphs again takes significantly longer than for SPADE and
CamFlow, probably because the graphs are larger.  For SPADE,
generalization of the \syscall{execve} benchmark takes much longer than for
other calls (though still only a few seconds).

The time required for the comparison stage is usually less than for
generalization (recall that in generalization we process two pairs of
graphs whereas in comparison we just compare the background and
foreground graphs).  Comparison of OPUS graphs again takes longest,
while comparison of CamFlow graphs again takes slightly longer than for
SPADE, perhaps because of the larger number of properties.  

In general, we can conclude that the time needed for ProvMark's
transformation, generalization, and comparison stages is acceptable
compared with the time needed for recording.  Most benchmarks complete
within a few minutes at most, though some that produce larger target graphs
take considerably longer; thus, at this point running all of the
benchmarks takes a few hours.  This seems like an acceptable
price to pay for increased confidence in these systems, and while
there is clearly room for
improvement, this compares favorably with
manual analysis, which is tedious and would take a skilled user
several hours.  In addition, we have monitored the memory usage and
found that ProvMark never used more than 75\% of memory on a 4GB
virtual machine, indicating memory was not a
bottleneck.

\subsection{Scalability}

\begin{figure}[tb]
\begin{tikzpicture}
\begin{axis}[
    xbar stacked,
    xlabel={Time (seconds)},
    symbolic y coords={scale1,scale2,scale4,scale8},
    xmin = 0,
    xmax = 1.5,
    y=-0.32cm,
	bar width=0.2cm,
	ytick=data,
    legend style={nodes={scale=0.75, transform shape}, legend columns=1}
    ]
\addplot+[xbar] plot coordinates {
		(.219,scale1)
		(.221,scale2)
		(.204,scale4)
        (.331,scale8)
};
\addplot+[xbar] plot coordinates {
		(.198,scale1)
		(.206,scale2)
		(.234,scale4)
        (.398,scale8)
};
\addplot+[xbar] plot coordinates {
		(.088,scale1)
		(.101,scale2)
		(.108,scale4)
      (.197,scale8)
};
\legend{Transformation,Generalization,Comparison}
\end{axis}
\end{tikzpicture}
\caption{Scalability results: SPADE+Graphviz}\label{chart:scalespade}
\bigskip
\begin{tikzpicture}
\begin{axis}[
    xbar stacked,
    xlabel={Time (seconds)},
    symbolic y coords={scale1,scale2,scale4,scale8},
    xmin = 0,
    xmax = 650,
    y=-0.32cm,
	bar width=0.2cm,
	ytick=data,
    legend style={nodes={scale=0.75, transform shape}, legend columns=1}
    ]
\addplot+[xbar] plot coordinates {
		(358.919,scale1)
		(358.965,scale2)
		(359.792,scale4)
        (364.211,scale8)
};
\addplot+[xbar] plot coordinates {
		(16.136,scale1)
		(16.341,scale2)
		(16.792,scale4)
        (18.716,scale8)
};
\addplot+[xbar] plot coordinates {
		(2.036,scale1)
		(2.139,scale2)
		(2.657,scale4)
        (3.702,scale8)
};
\legend{Transformation,Generalization,Comparison}
\end{axis}
\end{tikzpicture}
\caption{Scalability results: OPUS+Neo4J}\label{chart:scaleopus}
\bigskip
\begin{tikzpicture}
\begin{axis}[
    xbar stacked,
    xlabel={Time (seconds)},
    symbolic y coords={scale1,scale2,scale4,scale8},
    xmin = 0,
    xmax =6,
    y=-0.32cm,
	bar width=0.2cm,
	ytick=data,
    legend style={nodes={scale=0.75, transform shape}, legend columns=1}
    ]
\addplot+[xbar] plot coordinates {
		(.919,scale1)
		(.925,scale2)
		(1.007,scale4)
        (1.039,scale8)
};
\addplot+[xbar] plot coordinates {
		(.216,scale1)
		(.287,scale2)
		(.764,scale4)
        (2.240,scale8)
};
\addplot+[xbar] plot coordinates {
		(.128,scale1)
		(.130,scale2)
		(.234,scale4)
        (.369,scale8)
};
\legend{Transformation,Generalization,Comparison}
\end{axis}
\end{tikzpicture}
\caption{Scalability results: CamFlow+ProvJson}\label{chart:scalecamflow}
\end{figure}

Our experiments so far concentrated on benchmarking one syscall at a
time. The design of ProvMark allows arbitrary activity as the target
action, including sequences of syscalls. As mentioned above, we
can simply identify a target action
sequence using \texttt{\#ifdef TARGET} in order to let ProvMark generate background and foreground
programs respectively.

Of course, if the target activity consists of multiple syscalls, the
time needed for ProvMark to process results and solve the resulting
subgraph isomorphism problems will surely increase.  The generated
provenance graph will also increase in size and number of
elements. The NP-complete complexity of sub-graph isomorphism means
that in the worst case, the time needed to solve larger problem
instances could increase exponentially. This section investigates the
scalability of ProvMark when the size of the target action increases.

In figure \ref{chart:scalespade}-\ref{chart:scalecamflow}, the charts
show the time needed for the three processing subsystems of ProvMark in handling 
some scalability test cases on SPADE, OPUS and CamFlow respectively. The 
scalability test cases range from scale1 to
scale8. In test case scale1, the target action sequence is simply a
creation of a file and another deletion of the newly created file. In
test case scale2, scale4 and scale8, the same target action is repeated twice, 
four times, and eight times respectively.  The results show that the time
needed initially grows slowly for SPADE, but by scale8 the
time almost doubles compared to scale1.  For OPUS, the time increases
are dwarfed by the high overhead of transformation, which includes the
one-time Neo4j startup and access costs as discussed above.  For CamFlow, the
time needed approximately doubles at each scale factor.  Although further 
scalability experiments are needed to consider much larger graph or benchmark
sizes, these experiments do demonstrate that ProvMark can currently
handle short sequences of 10-20 syscalls without problems.

\subsection{Modularity and Extensibility}
ProvMark was designed with extensibility in mind.  Only the first two
stages (recording and generalization) depend on the details of the
recording system being benchmarked, so to support a new system, it
suffices to implement an additional recording module that uses the new
system to record provenance for a test executable, and (if needed) an
additional translation module that maps its results to Datalog.  

As discussed earlier, it was non-trivial to develop recording modules
for the three systems that produce reliable results.  In particular,
supporting CamFlow required several iterations and discussion with its
developers, which have resulted in changes to CamFlow to accommodate
the needs of benchmarking. This architecture has been refined through
experience with multiple versions of the SPADE and CamFlow systems,
and we have generally found the changes needed to ProvMark to maintain
compatibility with new versions to be minor.  Developing new
recording modules
ought to be straightforward for systems that work in a similar way to
one of the three currently supported.

If a provenance tool generates data in a format not currently
supported by ProvMark, an additional module for handling this type of
provenance data is needed. The need for new transformations should
decrease over time as more tools are introduced to ProvMark since
there are a limited number of common provenance formats.

As shown in Table~\ref{tab:modulesizes}, none of the three recording or transformation modules required more than 200 lines of code (Python 3 using only standard library imports).  We initially developed ProvMark with support for SPADE/DOT and OPUS/Neo4j combinations; thus, adding support for CamFlow and PROV-JSON only required around 330 lines of code.  
\begin{table}[tb]
  \centering
  \begin{tabular}{l|ccc}
    Module & SPADE & OPUS  & CamFlow \\
(Format) & (DOT) & (Neo4j) & (PROV-JSON) \\\hline
    Recording & 171 & 118 & 192 \\
Transformation & 74 & 122 & 128
  \end{tabular}
  \caption{Module sizes (Python lines of code)}
  \label{tab:modulesizes}
\end{table}
\normalsize

Finally, throughout the development period of ProvMark, the candidate
tools have all been updated and include new features and
behaviour. Over time, we have updated ProvMark to react to changes in
both SPADE and CamFlow, with few problems.

\subsection{Discussion and Limitations}\label{sec:limitations}
While we have argued that ProvMark is effective, fast enough
to produce timely results, and easy to extend, it also has some
limitations, which we now discuss.  

At this stage, ProvMark has only been applied to provenance tracking
at the operating-system level; in particular, for SPADE, OPUS and
CamFlow running under Linux.  We believe the same approach can be
adapted to other operating systems or other layers of distributed
systems, assuming an extension to deal with nondeterminstic or concurrent
activity as described below.

We have evaluated ProvMark using a small subset of individual
syscalls.  Creating benchmarks currently takes some manual effort and
it would be helpful to automate this process to ease the task of
checking all of the possible control flow paths each syscall can take,
or even generate multi-step sequences. In addition, the analysis and
comparison of the resulting benchmark graphs requires some manual effort and
understanding.

We have shown ProvMark is capable of handling benchmark 
programs with a small number of target system calls. Scaling up to
larger amounts of target activity or background activity appears
challenging, due to the NP-completeness and worst-case exponential
behavior of subgraph isomorphism testing.  However, for deterministic activity it
may be possible to take advantage of other structural characteristics of the
provenance graphs to speed up matching: for example, if matched nodes
are usually produced in the same order (according to timestamps), then
it may be possible to incrementally match the foreground and
background graphs.

Lastly, ProvMark currently handles deterministic activity only.
Nondeterminism (for example through concurrency) introduces additional
challenges: both the foreground and background programs might have
several graph structures corresponding to different schedules, and
there may be a large number of different possible schedules.  It also
appears neceessary to perform some kind of fingerprinting or graph
structure summarization to group the different possible graphs
according to schedule.  We may also need to run larger numbers of
trials and develop new ways to align the different structures so as to
obtain reliable results.  It appears challenging to ensure
completeness, that is, that all of the possible behaviors are
observed, especially since the number of possible schedules can grow
exponentially in the size of the program.
\section{Conclusions and Future Work}
\label{sec:concl}

This paper presents ProvMark, an automated approach to benchmarking
and testing the behavior of provenance capture systems.  To the best
of our knowledge it is the first work to address the unique challenges
of testing and comparing provenance systems.  We have outlined the 
design of ProvMark, and showed how it helps identify gaps in coverage,
idiosyncrasies and even a few bugs in three different provenance
capture systems.  We also showed that it has acceptable performance
despite the need to solve multiple NP-complete graph matching subproblems.
ProvMark is a significant step towards validation of such
systems and should be useful for developing correctness or
completeness criteria for them.

There are several directions for future work, to address the
limitations discussed in Section~\ref{sec:limitations}. First,
additional support for automating the process of creating new
benchmarks, or understanding and interpreting the results, would
increase the usefuless of the system, as would extending it to
provenance tracking at other levels, such as distributed system
coordination layers. Second, although we have evaluated scalability uo
to 10--20 target system calls, realistic applications such as
suspicious behavior analysis would require dealing with much larger
target activities and graphs. Finally, ProvMark cannot deal with
nondetermism, including concurrent I/O and network actvity, and
extending its approach to handle nondeterministic target activity is a
focus of current work.

\section*{Acknowledgments}
Effort sponsored by the Air Force Office of Scientific Research, Air
Force Material Command, USAF, under grant number
\grantnum{AFOSR}{FA8655-13-1-3006}. The U.S. Government and University
of Edinburgh are authorised to reproduce and distribute reprints for
their purposes notwithstanding any copyright notation thereon.  Cheney
was also supported by ERC Consolidator Grant Skye (grant number
\grantnum{ERC}{682315}).  This material is based upon work supported
by the Defense Advanced Research Projects Agency (DARPA) under
contract \grantnum{DARPA}{FA8650-15-C-7557} and the National Science Foundation under
Grant \grantnum{NSF}{ACI-1547467} and \grantnum{NSF}{SSI-1450277} (End-to-End Provenance). 
Any opinions, findings, and conclusions or
recommendations expressed in this material are those of the authors
and do not necessarily reflect the views of the National Science
Foundation.

We are grateful to Ajitha Rajan for comments on a draft of this paper
and to our shepherd Boris Koldehofe and anonymous reviewers for
insightful comments.


{ \bibliographystyle{ACM-Reference-Format}
\bibliography{paper}}

\appendix
\newpage
\section{ProvMark Documentation}

\par
ProvMark is a fully automated system that generates system level
provenance benchmarks for different provenance collection tools and systems, such as SPADE, OPUS and CamFlow. It is the main tool mentioned in the paper. This section provides documentation for how to use it and access the source code of ProvMark.

\subsection{ProvMark source and release page}

\par
The source code of ProvMark is stored in a Github repository and the
link is
\href{https://github.com/arthurscchan/ProvMark/}{here}\footnote{https://github.com/arthurscchan/ProvMark/}. ProvMark
is still undergoing development. There are many new updates after the
publication of this paper. In order for the reader of this paper to
use or test ProvMark with all the features mentioned in this paper, we
have provided a release version for this Middleware submission. The
link is
\href{https://github.com/arthurscchan/ProvMark/releases/tag/Middleware2019}{here}\footnote{https://github.com/arthurscchan/ProvMark/releases/tag/Middleware2019}. In
this release page, documentation on how to install, configure and use
the ProvMark tool has been provided. There is also a tarball with a
version of ProvMark that we are mentioning in this paper. In addition,
a set of sample results and additional experimental timing results is provided in this page.

\subsection{Directory structure of ProvMark source}
\begin{description}
\item[benchmarkProgram] Contains sample c programs for the collection of provenance information of different syscalls
\item[clingo] Contains the clingo code
\item[config] Contains the configuration profile of different tool choices for stage 1 and stage 2
\item[documentation] Contains the documentation for ProvMark
\item[genClingoGraph] Contains code to transform graph format
\item[processGraph] Contains code to handle graph comparison and generalization
\item[sampleResult] Contains sample benchmark result on our trial
\item[startTool] Contains tools to handle provenance collecting tools currently supported and retrieve result from them
\item[template] Contains html template for result generation
\item[vagrant] Contains vagrant files for those provenance collecting tools currently supported
\end{description}

\subsection{ProvMark Installation}

Installing ProvMark is simple, just clone the whole git repository. The current stable version corresponding to this paper is tagged by tag Middleware2019.
You could also directly download the source code tarball from the release page mentioned above.
\\
\footnotesize
\begin{verbatim}
git clone https://github.com/arthurscchan/ProvMark.git
cd ProvMark
git checkout Middleware2019
\end{verbatim}
\normalsize

\subsubsection{Vagrant installation}

In the vagrant folder, we have prepared the
\href{https://www.vagrantup.com/}{Vagrant}\footnote{https://www.vagrantup.com/}
script for the three provenance collecting tools currently
supported. If you have vagrant (v2.2.2 or greater) and virtual box
(v6.0 or greater) installed in your system, you can follow the steps
below to build up a virtual environment which everything (tools and
ProvMark) are installed.

\newpage

\subsubsection{Vagrant script for SPADE}

\footnotesize
\begin{verbatim}
cd ./vagrant/spade
vagrant plugin install vagrant-vbguest
vagrant up
vagrant ssh
\end{verbatim}
\normalsize

\subsubsection{Vagrant script for OPUS}

\footnotesize
\begin{verbatim}
cd ./vagrant/opus
vagrant plugin install vagrant-vbguest
vagrant up
vagrant ssh
\end{verbatim}
\normalsize
\par
\bigskip
To run OPUS, you also need a source or binary distribution for the OPUS system itself, which is available \href{https://github.com/DTG-FRESCO/opus}{here}\footnote{https://github.com/DTG-FRESCO/opus}


\subsubsection{Vagrant script for CamFlow}

\footnotesize
\begin{verbatim}
cd ./vagrant/camflowv045
vagrant plugin install vagrant-vbguest
vagrant up
vagrant halt
vagrant up
vagrant ssh
\end{verbatim}
\normalsize

It is necessary to reboot the VM (halt / up) so that the camflow-enabled kernel will be used.  This kernel should be highlighted as the first entry by the boot loader but if not, it should be selected.

After the above steps, you should be given a ssh connection to the
virtual machine from which you can start ProvMark on your chosen tools directly.
Note: the installation process can take an extended amount of time depending on your configuration.

\subsection{ProvMark configuration}

Configuration of ProvMark is done by modifying the \texttt{config/config.ini}
file. Changing this file should not be necessary in common cases.

In the ProvMark system, the first two stages collect provenance information from provenancee collecting tools and transform the provenance result into a clingo graph. Different tools need different process handling and type conversion. These configuration parameters are stored in the \texttt{config.ini} file.

Each profile starts with the name and includes three settings as follows:
\\
\begin{description}
\item[stage1tool] define the script (and parameter) for stage 1 handling when this profile is chosen

\item[stage2handler] define which graph handler is used to handle the raw provenance information when this profile is chosen

\item[filtergraphs] define if graph filtering is needed. The default value for SPADE and OPUS is false and true for CamFlow. The graph filtering is a mechanism provided by ProvMark to filter out obviously incomplete or incorrect graphs generated by the provenance systems. It can increase the accuracy for the resulting benchmark, but it will decrease the efficiency.
\end{description}
\par
\bigskip
Each profile defines one supporting tool and its configuration. If new tools are supported in ProvMark, a new profile will be created here.

\newpage

\subsection{ProvMark usage}

\subsubsection{Parameters}
\textbf{Currently Supported Tools:}
\begin{description}
\item[spg]    SPADE with Graphviz storage
\item[spn]    SPADE with Neo4j storage
\item[opu]    OPUS
\item[cam]    CamFlow
\end{description}
\bigskip
\textbf{Tools Base Directory:}\\
Base directory of the chosen tool, it is assumed that if you want to execute this benchmarking system on certain provenance collecting tools, you should have installed that tools with all dependencies required by the tools. If you build ProvMark with the given Vagrant script, the default tools base directory is shown as follows.
\\
\begin{description}
\item[SPADE] /home/vagrant/SPADE
\item[OPUS] /home/vagrant/opus (or where OPUS was manually installed)
\item[CamFlow] ./  (this parameter is ignored but some value must be provided)
\end{description}
\bigskip
\textbf{Benchmark Directory:}\\
Base directory of the benchmark program.\\
Point the script to the syscall choice for the benchmarking process. \\
\\ 
\textbf{Number of trials (Default: 2):}\\
Number of trials executed for each graph for generalization.\\
More trials will result in longer processing time, but provide a more accurate result as multiple trials can help to filter out uncertainty and unrelated elements and noise. \\
\\ 
\textbf{Result Type:}
\begin{description}
\item[rb] benchmark only
\item[rg] benchmark and generalized foreground and background graph only
\item[rh] html page displaying benchmark and generalized foreground and background graph
\end{description}

\subsubsection{Single Execution}
\par
This command will only call a specific benchmark program and generate benchmark with the chosen provenance system.
\\\\
Usage:
\scriptsize
\begin{verbatim}
./fullAutomation.py <Tools> <Tools Base Directory> <Benchmark Directory> [<Trial>]
\end{verbatim}
\normalsize
\bigskip
Sample:
\scriptsize
\begin{verbatim}
./fullAutomation.py cam ./ ./benchmarkProgram/baseSyscall/grpCreat/cmdCreat 2
\end{verbatim}
\normalsize

\subsubsection{Batch Execution}
\par
Automatically execute ProvMark for all syscalls currently supported. The runTests.sh script will search for all benchmark programs recursively in the default benchmarkProgarm folder and benchmark them one by one. It will also group the final result and post process them according to the given result type parameter.
\\\\
Usage:
\footnotesize
\begin{verbatim}
./runTests.sh <Tools> <Tools_Path> <Result Type>
\end{verbatim}
\normalsize
\bigskip
Sample:
\footnotesize
\begin{verbatim}
./runTests.sh spg /home/vagrant/SPADE rh
\end{verbatim}
\normalsize
\bigskip
For more examples of the usage please refer to the documentation provided on the release page or the README file at the parent directory of the source code.

\subsection{Sample Output}
\par
For the generation of the sample output, we have used the provided Vagrant script to build up the environment for the three provenance systems and execute a batch execution in each of the built virtual machine. The following command is used in each virtual machine respectively.

\subsubsection{SPADE}
\footnotesize
\begin{verbatim}
./runTests.sh spg /home/vagrant/SPADE rh
\end{verbatim}
\normalsize

\subsubsection{OPUS}
\footnotesize
\begin{verbatim}
./runTests.sh opu /home/vagrant/opus rh
\end{verbatim}
\normalsize

\subsubsection{CamFlow}
\footnotesize
\begin{verbatim}
./runTests.sh cam . rh 11
\end{verbatim}
\normalsize

The results on successful completion of this script are accessible in \texttt{finalResult/index.html}.

\subsubsection{Full Timing Result}
\par
From the data in Figures~\ref{chart:timedspade}--\ref{chart:timedcamflow} we can see the time needed for ProvMark to process some of the system calls using specific provenance systems. The full list of timing result containing all system calls included in the above batch execution can be found in the following path inside the source tarball from the release page. They are separated by the chosen provenance systems.
\\
\begin{description}
\item[SPADE] sampleResult/spade.time
\item[OPUS] sampleResult/opus.time
\item[CamFlow] sampleResult/camflow.time
\end{description}
\par
\bigskip
Each line in those result files representing one system call execution and all parameters in each line are separated by a comma. The first two parameters represent the provenance system and system calls chosen for this line of result. The remaining four floating-point numbers represent the time needed (in seconds) for each ProvMark subsystem to process that system call (with the chosen provenance system) in order. ProvMark will automatically time all the process and attached a line of timing result at the end of \texttt{/tmp/time.log} for each system call execution. You must have the privilege to write to that file to get the timing result.

\end{document}